\documentstyle[latexsym,amssymb,epsfig,aps,floats,eqsecnum,
preprint,amsfonts]{revtex}
\def\laq{\raise 0.4ex\hbox{$<$}\kern 
-0.8em\lower 0.62 ex\hbox{$\sim$}}
\def\gaq{\raise 0.4ex\hbox{$>$}\kern 
-0.7em\lower 0.62 ex\hbox{$\sim$}}
\def\H{{\cal H}}
\def\F{{\cal F}}
\def\r{\rho}
\def\a{\alpha}
\def\b{\beta}
\def\L{\Lambda}
\begin{document}
\draft
\bibliographystyle{unsrt}

\title{Static dilaton solutions and singularities 
in six dimensional warped compactification with higher derivatives}

\author{Massimo Giovannini\footnote{Electronic address: 
Massimo.Giovannini@ipt.unil.ch }}

\address{{\it Institute of Theoretical Physics, 
University of Lausanne}}
\address{{\it BSP-1015 Dorigny, Lausanne, Switzerland}}

\maketitle

\begin{abstract}
Static solutions with a bulk dilaton are derived in the context of six 
dimensional warped compactification. In the string frame, 
exponentially decreasing warp factors are identified 
with critical points of the low energy $\beta$-functions 
truncated at a given order in the string tension corrections. 
The stability of the critical points is discussed in the case 
of the first string tension correction. The singularity 
properties of the obtained solutions are analyzed and 
illustrative numerical examples are provided.
\end{abstract}
\vskip0.5pc
\centerline{Preprint Number: UNIL-IPT-23 October 2000}
\vskip0.5pc
\noindent
\newpage

\renewcommand{\theequation}{1.\arabic{equation}}
\setcounter{equation}{0}
\section{Formulation of the problem} 
Consider a $(4\,+2)-$dimensional space-time (consistent 
with four dimensional Poincar\'e invariance) of the form \cite{m1,m2} 
\begin{equation}
ds^2 =g_{\mu\nu} dx^{\mu} dx^{\nu} = 
\sigma(\rho) G_{a b} dx^a ~d x^b - d\rho^2 - \gamma(\rho) 
d\theta^2,
\label{line}
\end{equation}
where the (global) behavior of 
$\sigma(\rho)$ and $\gamma(\rho)$
 should be determined by consistency 
with the appropriate equations of motion following, for 
instance, from the 
$(4\,+2)$-dimensional Einstein-Hilbert action 
\footnote{The conventions 
of the present paper are the following : the signature 
of the metric is mostly minus, Latin indices run over 
the $(3+1)$-dimensional space, whereas Greek indices run 
over the whole $(4+2)$-dimensional space.}. 
This type of compactification differs  from ordinary Kaluza-Klein
schemes \cite{m1,m2} (see also \cite{ak,viss}). 
The form of the line element given in Eq. (\ref{line}) 
can be generalized to higher 
dimensions \cite{ran}, although, for the present investigation, 
a six-dimensional metric will be considered.

Most of the studies dealing with six-dimensional warped compactification
assume that the underlying 
theory of gravity is of Einstein-Hilbert type.
Solutions with exponentially decreasing warp factors 
have been obtained in the presence of a bulk 
cosmological constant \cite{ck,cp} supplemented 
by either global \cite{vil,greg} or local \cite{gs} 
string-like defects (see also \cite{s1,s2}).
The shape of the warp factor far and away
from the core of the defect is always determined, in the quoted 
examples, according to the six-dimensional Einsten-Hilbert 
description.
Six-dimensional warped compactifications represent a 
useful example also because they are a non-trivial generalization 
of warped compactifications in five dimensions \cite{RS}.
 
Up to now the compatibility of warped compactification  
with theories different from the Einstein-Hilbert description have 
been analyzed mainly in the five-dimensional case. In particular 
the attention has been payed to gravity theories with higher derivatives 
\cite{hd,hda,hdb}. In \cite{rm}
 the interesting problem of the compatibility of a five-dimensional 
compactification scheme with higher derivative gravity and dilaton 
field has been investigated. 
 In \cite{mg} the simultaneous presence 
of higher dimensional hedgehogs and of higher derivatives 
gravity theories has been discussed in a seven-dimensional space-time.

The simultaneous 
presence of dilaton field and higher order curvature 
corrections is relevant for 
different reasons. There are two possible approaches 
to warped compactification schemes. In the first 
approach it is
assumed, from the very beginning, that 
warped compactifications have nothing to do with 
string theory.  The problem then reduces to the search of a 
suitable field theoretical model of higher dimensional (global or local)
defects living along the transverse dimensions.
A complementary approach is to 
postulate that warped compactifications are 
 (somehow) connected to string theory. 
Even in this second approach, however, the 
dilaton field and the (possible) higher curvature terms 
(arising from the string tension corrections) are usually
neglected \cite{RS}. The assumption that the dilaton 
field is strictly constant and that the string tension 
corrections are vanishing should be relaxed in 
order to investigate the (possible) interplay of warped 
compactification schemes with string models. 

Five-dimensional warped compactifications have been studied
in the context of generalized gravity theories 
with higher derivatives \cite{hd,hda,hdb}. The 
inclusion of the dilaton field poses, however some 
problem. The main results concerning the 
interplay between 
 five-dimensional warped compactifications 
and string motivated effective action can be 
summarized as follows \cite{rm}. If the dilaton is constant 
(but higher order curvature corrections are still present) 
five-dimensional warped compactifications 
are consistent with the effective action derived 
from the string amplitudes corrected to first 
order in the string tension. 

If the dilaton is linear 
in the bulk coordinate warped compactification is 
not consistent with the presence of string tension corrections unless
the conformal coupling between the dilaton and the 
string tension corrections is absent \footnote{ The effective 
action used in order to derive these results has been studied
Einstein frame.} \cite{rm}. 
Interesting solutions interpolating between a naked singularity
and a warped regime (for large bulk coordinate) have also been obtained. 
The purpose of the present investigation is to 
is to analyze the solutions of the metric (\ref{line})
in the context 
of the string effective action \cite{steff1,steff2} with and without 
higher derivatives corrections \cite{hd1,hd2} (see also \cite{hd3,hd4}). 

Up to now the compatibility of six-dimensional 
warped compactification 
with gravity theories inspired by string amplitudes 
has not been analyzed. In 
six-dimensional warped compactifications
the transverse space is not flat (in contrast with the  
five-dimensional case). This observation implies on one 
hand that the internal space is larger and, on the other 
hand, that two warp factors are generically 
allowed.
The tree-level solutions have more general singularity 
properties if compared to the five-dimensional case.  
These solutions 
are ``Kasner-like'' and they are the static 
analog of their time-dependent counterpart which is often 
discussed in cosmological solutions \cite{gv}.
The inclusion of string tension corrections 
introduces also differences with respect to the 
five-dimensional case. 

In  six dimensions exponentially 
decreasing warp factors correspond to critical points 
of the $\beta$ functions computed at a given order 
in the string tension. 
Define, in fact,   ${\cal H} = \partial_{\rho}(\ln\sigma)$ and 
${\cal F} = \partial_{\rho}(\ln\gamma)$. By critical 
points we mean those (stable or unstable) 
solutions  for which $ {\cal H}$ and ${\cal F}$ are 
simultaneously constant and negative. 
 The critical points 
of the system correspond to  a static dilaton fields 
which either increases or decreases (linearly) 
for large $\r$. Also the constant dilaton solution 
is possible. This situation should be contrasted with 
the five-dimensional case where only one warp factor 
is present.

The critical points  are 
not always stable. If the initial conditions of 
the dilaton and of the warp factors are given around a given 
critical point (say for $\rho =\rho_0$) it can happen 
that for larger $\rho$   the compatibility with the 
$\beta$-functions will drive the solution away from the (original)
critical point. In this 
case singularity may also be developed.

The present analysis is not meant to be exhaustive and suffers 
of two obvious limitations. In order to make an explicit 
calculation  the effects 
of the first string tension correction has been studied. 
This is just an example since, when
singularities are developed, {\em all } the string tension corrections should 
be included.  The second point to be emphasized is that 
possible corrections in the dilaton coupling have also been 
neglected. This might be justified in some regimes 
of the solutions but it is not justified in more general
terms.  With these two warnings in mind the reported 
results should be understood more as possible indications 
than as a firm conclusion. 

The present analysis has been performed in the string frame where 
the dilaton field is directly coupled both to the Einstein-Hilbert 
term and to the first string tension corrections. 
It is interesting to notice that singularity properties 
of a given solution may change from one frame to the other. A linearly 
increasing dilaton (in the string frame) results in curvature 
singularities in the Einstein frame (and vice-versa). 
The Einstein and the string frame are, however, physically equivalent. 
Suppose that the  curvature invariants are regular in one frame 
and suppose that the dilaton is a smooth function 
of the bulk radius (interpolating, for instance, between 
two constant regimes). Then, the geometry looks regular in both frames 
enforcing the physical equivalence of the two descriptions. 

In order to assess that six-dimensional warped 
compactification is fully consistent with the presence 
of the dilaton field and of string tension corrections some 
requirements should be, in our opinion, satisfied.
The goal would be to obtain stable warped solutions 
with constant dilaton. Stable warped solutions means 
that the critical point of the low energy $\beta$ functions
(truncated to first order in the $\alpha'$ corrections) lead to exponentially 
decreasing warp factors. The constancy of the dilaton field $\phi$ 
implies the constancy of the coupling constant (i.e. $\exp{[\phi/2]}$) 
and this is a necessary (even though not sufficient) condition 
for the stability of the solutions. Notice that 
once the dilaton field and the string tension corrections 
are simultaneously present the problem of the singularity of the geometry 
cannot be disentangled from the problem of the dilaton relaxation.

The plan of the present investigation is then the following.
In Section II the tree-level solution of the low energy string 
effective action will be studied. In Section III the first 
string tension correction will be included and the corresponding 
equations of motion will be derived. In Section IV 
the critical points of the obtained dynamical system 
will be analyzed. In Section V the stability of the critical 
points will be scrutinized and some numerical examples will be presented. 
Section VI contains some concluding remarks. In the Appendix 
useful technical results are reported.

\renewcommand{\theequation}{2.\arabic{equation}}
\setcounter{equation}{0}
\section{Tree-level solutions} 
If the curvature of the background is sufficiently small 
in units 
of the string length 
\footnote{ In the discussion string units (i.e. $\lambda_s =1$) will 
be often used.} (denoted by $\lambda_s$) the massless modes 
of the string are weakly coupled and the dynamics 
can be described by  
the string effective action in six dimensions \cite{steff1,steff2}
\begin{equation}
S^{(0)} = \int d^6 x \sqrt{ - g} {\cal L}^{(0)} \equiv 
 - \frac{1}{2\lambda_s^4} \int d^6 x \sqrt{- g} e^{-\phi} 
\biggl[ R 
+ g^{\alpha\beta} \partial_{\alpha} \phi \partial_{\beta}
\phi + \Lambda  \biggr],
\label{action1}
\end{equation}
where $\phi$ is the dilaton field, $\lambda_{s} = 
\sqrt{\alpha'} $ ($\alpha'$ being the string tension). 
Eq. (\ref{action1}) is written in the string frame where 
the string scale is constant and the Planck scale depends upon 
the value of the dilaton coupling, i.e. 
$g(\phi) = e^{\phi/2}$. If the dilaton coupling is constant 
the the string frame coincides with Einstein frame. 
If, as in the present case, 
$\phi = \phi(\rho)$ the two frames are equivalent up to a conformal 
transformation. In the present paper the string frame will be used.
In Eq. (\ref{action1}) the minimal field 
content (i.e. graviton and dilaton) has been assumed together with a
bulk cosmological constant $\L$.
The effective action (\ref{action1}) is derived by requiring 
that the usual string scattering amplitudes are 
correctly reproduced to the lowest order in $\alpha'$. 
The requirement that the equations of motion derived 
from Eq. (\ref{action1}) are satisfied 
in the metric (\ref{line}) 
is equivalent to the requirement that the background is 
conformally invariant to the lowest order in $\alpha'$. 
Notice that in eq. (\ref{action1}) the contribution of the antisymmetric 
tensor field has been neglected. This is justified within the spirit 
of the present analysis but this might not be justified in
more general terms. We will come back on this point in the 
following Sections.

The  equations of motion derived from the action of Eq. (\ref{action1}) 
can be written as 
\begin{eqnarray}
&& R - g^{\alpha\beta} \partial_{\alpha} \phi \partial_{\beta} \phi + 2
g^{\alpha\beta} \nabla_{\alpha} \nabla_{\beta} \phi + \Lambda =0,
\label{beta1}\\
&& R_{\mu\nu} - \frac{1}{2} g_{\mu\nu} R 
+ \frac{1}{2} g_{\mu\nu} g^{\alpha\beta} \partial_{\alpha} \phi
 \partial_{\beta}\phi
- g_{\mu\nu} g^{\alpha\beta} \nabla_{\alpha} \nabla_{\beta} \phi
+ \nabla_{\mu} \nabla_{\nu} \phi - 
\frac{\Lambda}{2} g_{\mu\nu} =0.
\label{beta2}
\end{eqnarray}
By now using the metric of Eq. (\ref{line})  the explicit form  of the 
 $(a,\,a)$ and $(\theta,\,\theta)$ components
of Eq. (\ref{beta2}) will be  
\footnote{As previously mentioned, for convenience, the 
following notations are used ${\cal H} = (\ln{\sigma})'$, 
${\cal F} = (\ln{\gamma})'$ 
where $' = \partial_{\rho}$.}
\begin{eqnarray}
&& 2\,{\L} + {\F}^2 + 3\,\F\,\H + 6\,{\H}^2 + 2\,\F' + 6\,\H' - 
  2\,\F\,{\phi}' - 6\,\H\,{\phi}' + 2\,{{\phi}'}^2 - 
  4\,{\phi}'' =0,
\label{betaa}\\
&& 2\,{\L} + 10\,{\H}^2 + 8\,\H' - 8\,\H\,{\phi}' + 
  2\,{{\phi}'}^2 - 4\,{\phi}''=0,
\label{betab}
\end{eqnarray}
whereas the explicit form of the  of Eq. (\ref{beta1}) will be
\begin{equation}
 -2\,{\L} - {\F}^2 - 4\,\F\,\H - 10\,{\H}^2 - 2\,\F' - 8\,\H' + 
  2\,\F\,{\phi}' + 8\,\H\,{\phi}' - 2\,{{\phi}'}^2 + 
  4\,{\phi}'' =0.
\label{betac}
\end{equation}
Notice that Eq. (\ref{beta1}) has been multiplied by 
a factor two and Eq. (\ref{beta2}) has been multiplied by a factor of four
in order to get rid of rational coefficients.
In deriving Eqs. (\ref{betaa})--(\ref{betac}) it has been assumed that, in Eq. 
(\ref{line}) $G_{a\,b} \equiv \eta_{a\,b}$ where $\eta_{a\,b}$ is the 
Minkowski metric.  
Eqs. (\ref{betaa})--(\ref{betac}) admit exact solutions whose 
explicit form can be written as:
\begin{eqnarray}
&&\sigma(\rho) = \tanh{\biggl[ \frac{\sqrt{ - \lambda}}{2} ( \rho - \rho_0)
\biggr]}^{ 2 \alpha} ,
\label{tlev1}\\
&& \gamma(\rho) = \tanh{\biggl[ \frac{\sqrt{ - \lambda}}{2} ( \rho - \rho_0)
\biggr]}^{ 2 \beta},
\label{tlev2}\\
&& e^{\phi(\rho)} = \frac{1}{2} \biggl\{ \sinh{\biggl[ \frac{ 
\sqrt{- \Lambda}}{2} (\rho - \rho_0) \biggr]}\biggr\}^{ 4 \alpha + \beta -1}
     \biggl\{ \cosh{\biggl[ \frac{ 
\sqrt{- \Lambda}}{2} (\rho - \rho_0) \biggr]}\biggr\}^{ -4 \alpha - \beta -1},
\label{tlev}
\end{eqnarray}
The exponents $\alpha$ and $\beta$
satisfy the condition $ 4 \alpha^2 + \beta^2 =1$. 
Eqs. (\ref{tlev1})--(\ref{tlev}) lead to a physical singularity 
for $\r \rightarrow \r_0$. 
All curvature invariants associated with the 
metric (\ref{line}) for the specific solution given in Eqs. 
(\ref{tlev1})--(\ref{tlev}) diverge 
for $\rho\rightarrow \r_0$ (see Appendix A for the details). If $
\alpha = \beta$ the Weyl invariant vanishes but the other invariants are 
still singular.
 
If $\Lambda \rightarrow 0$, Eqs. (\ref{betaa})--(\ref{betac}) 
are solved by 
\begin{eqnarray}
&&\sigma(\rho) = (\r - \r_{0})^{2 \alpha}, 
\label{k1}\\
&& \gamma(\rho) = (\r - \r_0)^{2 \beta},
\label{k2}\\
&& e^{\phi(\rho)} = (\r - \r_0)^{ -1 + 4 \alpha + \beta},
\label{k3}
\end{eqnarray}
provided $ 4 \alpha^2 + \beta^2 =1$.
Eqs. (\ref{tlev1})--(\ref{tlev}) and (\ref{k1})--(\ref{k3})
are  Kasner-like \footnote{ For truly Kasner
solutions  the sum of the exponents (and of their {\em squares}) 
has to equal one. In the present case only the sum 
of the squares is constrained and this is the reason why 
these solutions are often named ``Kasner-like''.} 
solutions whose time-dependent 
analog has been widely 
exploited in the context of string cosmological solutions 
\cite{gv}.

In the case of $\Lambda <0$ this system of equation has a further 
non trivial solution which is given by ${\cal H}$, ${\cal F}$ and 
$\phi'$ all constant. In this case the solution of the 
previous system of equations is given by 
\begin{eqnarray}
&& 4 \Lambda + 4 {\cal H}^2 + {\cal F}^2 =0, \,\,\,\, 2 \phi' = 4 {\cal H} 
+ {\cal F}.
\label{trecrit}
\end{eqnarray}
In the particular case where $\sigma(\rho) \propto \gamma(\rho) $ (
i.e. ${\cal H} = {\cal F}$), the solution is ${\cal H} = - \sqrt{ - 4\L/5}$.
As we will discuss in Section IV this solution is not always 
stable.

The presence of curvature singularities in the tree-level 
solutions obtained in Eqs. (\ref{tlev1})--(\ref{tlev}) and 
(\ref{k1})--(\ref{k2}) suggests that there are physical regimes where 
the curvature of the geometry will approach the 
string curvature scale. In this 
situation higher order (curvature) corrections may play a role 
in stabilizing the solution and should be considered. 
The following part of this investigation will then deal 
with the inclusion of the first $\alpha'$ correction. This
analysis is of course not conclusive {\em per se} since 
also higher orders in $\alpha'$ should be considered as it has 
been argued in Section I. 
\renewcommand{\theequation}{3.\arabic{equation}}
\setcounter{equation}{0}
\section{First order $\alpha'$ corrections} 
Consider now the first $\alpha'$  correction to the action $S^{(0)}$ 
presented in Eq. (\ref{action1}). The full action is, in this case
\cite{hd1}
\begin{equation}
S= S^{(0)}\, +\, S^{(1)} = -  \frac{1}{2\lambda_s^4} \int\, d^6 x\, 
\sqrt{- g}\, e^{-\phi}\, \biggl[ R\, 
+\, g^{\alpha\beta} \partial_{\alpha} \phi \partial_{\beta}\phi \,
+\, \Lambda  - \epsilon \,R_{\mu\nu\alpha\beta}\, R^{\mu\nu\alpha\beta} 
\biggr],
\label{eff1}
\end{equation}
where $ \epsilon = k \alpha'/4$  and the constant $k$ takes different values 
depending upon the specific theory ( $ k =1$ for the bosonic theory,
 $k= 1/2$  for the heterotic theory).
Notice that the assumption made in the previous 
Section (concerning the absence of antisymmetric 
tensor field) reflects in the fact that the first $\alpha'$ correction 
appears precisely in the form reported in eq. (\ref{eff1}). 
If the antisymmetric tensor field would not be vanishing 
the first string tension correction to the 
tree-level action would look like \cite{hd1}
\begin{eqnarray}
&&-\epsilon\biggl( R_{\mu\nu\alpha\beta} R^{\mu\nu\alpha\beta} - \frac{1}{2} 
R^{\mu\nu\alpha\beta} H_{\alpha\beta\lambda} H^{\lambda}_{\,\mu\nu} \biggr.
\nonumber\\
&&+ 
\biggl.\frac{1}{24} H_{\mu\nu\lambda} H^{\nu}_{\,\rho\alpha} H^{\rho\sigma\lambda} 
 H_{\sigma}^{\,\mu\alpha} - \frac{1}{8} H_{\mu\alpha\beta} H_{\nu}^{\,\alpha\beta}  
H^{\mu\rho\sigma} H^{\nu}_{\,\rho\sigma}\biggr) + {\cal O}(\epsilon^2)
\label{eff2}
\end{eqnarray}
where $H_{\mu\nu\alpha}$ is the antisymmetric tensor field strength. 
If $H_{\mu\nu\alpha}$ vanishes identically the only correction to the tree-level action 
is the one coming from the first term of Eq. (\ref{eff2}). However, if 
$H_{\mu\nu\alpha}\neq 0$, the situation changes qualitatively. We leave 
this intriguing issue for future investigations.

The fields appearing in the action (\ref{eff1}) 
can be redefined (preserving
the perturbative string amplitudes) \cite{hd1,hd3}.
In order to discuss actual solutions it is useful to perform a field 
redefinition (keeping the $\sigma$ model parameterization of the action) 
that 
eliminates terms with higher than second derivatives from the effective 
equations of motion \cite{hd1,hd2} (see also \cite{hd3,hd4}). 
In six dimensions the field redefinition can be written as
\begin{eqnarray}
&& \overline{g}_{\mu\nu} = g_{\mu\nu} + 16 \epsilon \biggl[ R_{\mu\nu}\, 
-\, \partial_{\mu}\phi \partial_{\nu} \phi \,+\, 
g_{\mu\nu} g^{\alpha\beta} \partial_{\alpha} 
\phi \partial_{\beta} \phi \biggr],
\nonumber\\
&& \overline{\phi} = \phi + \epsilon\, 
\biggl[ R + (3 + 2 n) g^{\alpha\beta}g^{\rho\sigma} 
 \partial_{\alpha}\phi\, \partial_{\beta} \phi\, 
\partial_{\rho}\phi\,\partial_{\sigma}\phi\, \biggr],
\end{eqnarray}
where $n$ (the number of transverse dimensions)  is equal to 2 
in the case of the present analysis.
Dropping the bar in the redefined fields the action reads :
\begin{equation}
S = - \frac{1}{2\lambda_s^4} \int d^6 x \sqrt{- g} e^{-\phi} \biggl[ R 
+ g^{\alpha\beta} \partial_{\alpha} \phi \partial_{\beta}\phi + \Lambda 
 - \epsilon\biggl( R_{\rm EGB}^2 - g^{\alpha\beta}
g^{\rho\sigma} 
\partial_{\alpha} \phi\partial_{\beta} \phi \partial_{\rho} \phi 
\partial_{\sigma} \phi
\biggl)\biggr],
\label{action2}
\end{equation}
where 
\begin{equation}
R^2_{\rm EGB} = R_{\mu\nu\alpha\beta}R^{\mu\nu\alpha\beta} - 4 R_{\mu\nu} 
R^{\mu\nu} + R^2,
\end{equation}
is the Euler-Gauss-Bonnet invariant (which coincides, in four space-time 
dimensions, with the Euler invariant \footnote{The Euler-Gauss-Bonnet 
invariant is particularly useful in order to parameterize quadratic corrections 
in higher dimensional cosmological models \cite{madore,mgg,rt}.}). 
The equations of motion 
can be easily derived by varying the action (\ref{action2}) 
 with respect to the 
metric and with  respect to the 
dilaton field (see Appendix B for details). The result is that 
\begin{eqnarray}
&&2\,{\L} + {\F}^2 + 3\,\F\,\H + 6\,{\H}^2 + 2\,\F' + 6\,\H' - 
  2\,\F\,{\phi}' - 6\,\H\,{\phi}' + 2\,{{\phi}'}^2 - 
  4\,{\phi}'' 
\nonumber\\
&&+ \epsilon\,\bigl[ -3\,{\F}^2\,{\H}^2 - 9\,\F\,{\H}^3 - 3\,{\H}^4 - 
     6\,{\H}^2\,\F' - 12\,\F\,\H\,\H' - 6\,{\H}^2\,\H' + \bigr. 
\nonumber\\     
&& \bigl.6\,{\F}^2\,\H\,{\phi}' + 24\,\F\,{\H}^2\,{\phi}' + 
     18\,{\H}^3\,{\phi}' + 12\,\H\,\F'\,{\phi}'\bigr. 
\nonumber\\
&&\bigr. + 12\,\F\,\H'\,{\phi}' + 24\,\H\,\H'\,{\phi}' - 
     12\,\F\,\H\,{{\phi}'}^2 - 12\,{\H}^2\,{{\phi}'}^2\bigr. + 
\nonumber\\
&&     \bigl.2\,{{\phi}'}^4 + 12\,\F\,\H\,{\phi}'' + 
     12\,{\H}^2\,{\phi}'' \bigr] =0,
\label{uno}\\
&&2\,{\L} + 10\,{\H}^2 + 8\,\H' - 8\,\H\,{\phi}' + 
  2\,{{\phi}'}^2 - 4\,{\phi}'' +
\nonumber\\ 
&&  \epsilon\,\bigl[ -15\,{\H}^4 - 24\,{\H}^2\,\H' + 48\,{\H}^3\,{\phi}' + 
     48\,\H\,\H'\,{\phi}' - 24\,{\H}^2\,{{\phi}'}^2 + \bigr.
\nonumber\\ 
&&     \bigl.2\,{{\phi}'}^4 + 24\,{\H}^2\,{\phi}'' \bigr] =0,
\label{due}\\
&&-2\,{\L} - {\F}^2 - 4\,\F\,\H - 10\,{\H}^2 - 2\,\F' - 8\,\H' + 
  2\,\F\,{\phi}' + 8\,\H\,{\phi}' - 2\,{{\phi}'}^2 + 
  4\,{\phi}'' 
\nonumber\\
&& + \epsilon\,\bigl[ 6\,{\F}^2\,{\H}^2 + 24\,\F\,{\H}^3 + 15\,{\H}^4 + 
     12\,{\H}^2\,\F' + 24\,\F\,\H\,\H' + 24\,{\H}^2\,\H' \bigr.
\nonumber\\
&& \bigl.- 4\,\F\,{{\phi}'}^3 - 
     16\,\H\,{{\phi}'}^3 + 6\,{{\phi}'}^4 - 
     24\,{{\phi}'}^2\,{\phi}'' \bigr] =0.
\label{tre}
\end{eqnarray}
In the limit $\epsilon\rightarrow 0$ the equations 
derived in Section II are recovered.
Eqs. (\ref{uno})--(\ref{tre}) can be studied in order to investigate 
two separate issues. The first one is the 
existence of critical points leading to exponentially decreasing warp factors. 
The second issue would be to analyze the stability of these 
critical points. Indeed, in the solutions given in Eqs. 
(\ref{tlev1})--(\ref{tlev}) a singularity is developed. This result 
is based, however, on the tree-level action. If the first
$\alpha'$ correction is included this feature might change.
Moreover, if $\Lambda =0$ there are no warped solutions at 
tree-level. These solutions might emerge, however, when $\alpha'$ 
corrections are included. 
For instance, the  ``Kasner-like'' branch of the solution  might 
be analytically connected to a ``warped'' regime where ${\cal H}$ and 
${\cal F}$ are both constant and negative. These questions will be the subject
of the following Section.

In closing this section we want to recall that in \cite{malnun} 
it has been claimed that (five-dimensional) warped compactifications 
cannot be obtained in the low energy supergravities 
coming from string theory. In this context it has also been suggested 
that higher order corrections to the Einstenian action might help in evading 
this conclusion. In the following we will explore an analogous possibility.
 
\renewcommand{\theequation}{4.\arabic{equation}}
\setcounter{equation}{0}
\section{Dynamical system and critical points} 
Defining
\begin{equation}
x(\r) \equiv {\cal H}(\r),~~~~y(\r) \equiv {\cal F}(\r),~~~~~~
z(\r) \equiv \phi'(\r) ,
\end{equation}
 Eqs. (\ref{uno})--(\ref{tre})  can be written as
\begin{eqnarray}
&&p_{x}(\r)\, x' + p_y(\r)\, y' + p_z(\r)\, z' + p_0(\r) =0,
\label{ss1}\\
&& q_{x}(\r)\, x' +  q_z(\r) z' + q_0(\r) =0,
\label{ss2}\\
&& w_x(\r) \, x' + w_y(\r)\, y' + w_z(\r)\,z' + w_0(\r) =0.
\label{sys1}
\end{eqnarray}
The $p$ are 
\begin{eqnarray}
&& p_{x}(\r) = 6 \{ 1 - \epsilon [ x ( x + 2 y) - 2 z ( y + 2 x)]\},\,\,\,\,
\nonumber\\
&& p_{y}(\r) = 2 [ 1 - 3 \epsilon x ( x - 2 z)],\,\,\,\,\,
 p_z(\r) = - 4 [ 1 - 3 \epsilon x ( x + y)],
\nonumber\\
&& p_0(\r) = 2 \L + y ( y + 3 x) + 6 x^2 - 2 z ( y + 3 x) + 2 z^2
\nonumber\\
&& +\epsilon[ - 3 x^2 ( y^2 + 3 x y + x^2) + 6 x  ( x + y) ( y + 3 x) z
-12 x ( x + y) z^2 + 2 z^4],
\end{eqnarray}
whereas the $q$ are 
\begin{eqnarray}
&& q_x(\r) = 8 [ 1 - 3 \epsilon x ( x - 2 z)],\,\,\,\,\,
 q_z(\r) = - 4 ( 1 - 6 \epsilon x^2),
\nonumber\\
&& q_0(\r) = 2 \L + 2 ( z^2 - 4 x z + 5 x^2) +
\epsilon [ - 15 x^4 + 48 x^3 z - 24 x^2 z^2  + 2 z^4],
\end{eqnarray}
and finally the $w$ are
\begin{eqnarray} 
&& w_x(\r) = - 8 [ 1 - 3 \epsilon x ( x + y)],
\nonumber\\
&& w_y(\r) = - 2 [ 1 - 6 \epsilon x^2],\,\,\,\,\, 
w_z(\r) = 4 ( 1 - 6 \epsilon z^2),
\nonumber\\
&& w_0(\r) = - 2 \L - ( y^2 + 4 x y + 10 x^2 ) + 2 ( y + 4 x) z - 2 z^2 
\nonumber\\
&& + \epsilon [ 3 x^2 ( 2 y^2  + 8 x y + 5 x^2) - 4 ( y + 4 x) z^3 + 6 z^4].
\end{eqnarray}
Eqs. (\ref{ss1})--(\ref{sys1}) 
can be analyzed both analytically 
and numerically. Interesting analytical conclusions can be obtained, for 
instance, in the case $x(\rho)= y(\rho)$. In the remaining 
part of the present Section the stability of the system will be analyzed. 
Physical (i.e. curvature) singularities will be investigated. 
The logic will be, in short, the following. The critical 
points will be firstly studied analytically. Indeed, it happens 
that, in spite of the apparent complications of the system, there 
are limits in which analytical expressions can be given. Numerical 
examples will be given for the other cases which are not solvable 
analytically. The examples studied in the present Session will be 
exploited in Section V for some explicit numerical solutions.

The critical points of the system derived in Eqs. 
(\ref{ss1})--(\ref{sys1}) are defined
to be the one for which \cite{dyn}
\begin{equation}
 x'(\r)=0,\,\,\, y'(\r) =0,\,\,\,z'(\r) =0.
\label{defcrit}
\end{equation}
The critical points interesting for warped compactifications 
are the ones for which $ x(\r)$ and  $y(\r)$ are not only constant 
but also negative. These points lead, for large $\rho$, 
to exponentially decreasing warp factors in the six-dimensional
metric.  

In the case of Eq. (\ref{defcrit}) the solution of the system 
reduces to a system of three algebraic equations in the unknowns 
$(x,y,z)$. The three equations of the system are 
simply given by the conditions 
\begin{equation}
p_0(\r) =0,\,\,\,\, q_0(\r) =0,\,\,\,\, w_0(\r) =0,
\end{equation}
namely, 
\begin{eqnarray}
&& 2 \L + y ( y + 3 x) + 6 x^2 - 2 z ( y + 3 x) + 2 z^2
\nonumber\\
&& +\epsilon[ - 3 x^2 ( y^2 + 3 x y + x^2) + 6 x  ( x + y) ( y + 3 x) z
-12 x ( x + y) z^2 + 2 z^4] =0,
\label{prim}\\
&& 2 \L + 2 ( z^2 - 4 x z + 5 x^2) +
\epsilon [ - 15 x^4 + 48 x^3 z - 24 x^2 z^2  + 2 z^4]=0,
\label{sec}\\
&& - 2 \L - ( y^2 + 4 x y + 10 x^2 ) + 2 ( y + 4 x) z - 2 z^2 
\nonumber\\
&&
+ \epsilon [ 3 x^2 ( 2 y^2  + 8 x y + 5 x^2) - 4 ( y + 4 x) z^3 + 6 z^4]=0.
\label{ter}
\end{eqnarray}
By summing up, respectively, Eqs. (\ref{prim}) 
and (\ref{sec}) with Eq. (\ref{ter}) 
the following two algebraic conditions are obtained:
\begin{eqnarray}
&& ( 4 x + y - 2 z)[ 3 \epsilon x^3 - 4 \epsilon z^3 + 
3 \epsilon x^2( y + 2 z) + x ( -1 + 6 \epsilon y z)]=0,
\nonumber\\
&& ( 4 x + y - 2 z) 
[ y ( -1 + 6 \epsilon x^2) - 4 \epsilon z ( z^2 - 3 x^2)]=0.
\label{rel1}
\end{eqnarray}
A third algebraic relations is 
obtained  by subtracting Eq. (\ref{sec}) from Eq. (\ref{prim}):
\begin{equation}
( x - y) ( 4 x + y - 2 z) [ -1 + 3 \epsilon x ( x - 2 z)] =0.
\label{rel2}
\end{equation}
If $x=y$  Eqs. (\ref{prim})--(\ref{ter}) are solved {\em provided} 
\begin{eqnarray}
&& x = y , \,\,\,\,\, z = \frac{ 4 x  + y}{2}, \,\,\,\,\,
\label{alge0}\\
&&16 \L + 20 x^2 + 265 \epsilon x^4 =0,
\label{alge}
\end{eqnarray}
and an explicit solution corresponding to the critical points of the system 
(\ref{prim})--(\ref{ter}) can then be written as 
\begin{eqnarray}
&& \sigma(\rho) = e^{ \bigl(\frac{\rho}{\rho_1}\bigr)}, \,\,\,\,\, 
\gamma(\rho) = \gamma_0 \sigma(\rho) 
\nonumber\\
&& \phi(\rho) = \frac{5}{2} \bigl(\frac{\rho}{\rho_1}\bigr) + \phi_0
\end{eqnarray}
where $\gamma_0$ and $\phi_0$ are integration constants and where 
\begin{equation}
\rho_1 = \pm \sqrt{-\frac{5}{8 \L}}
\sqrt{ 1 \mp \sqrt{ 1 - \frac{212}{5} \L\epsilon}}
\label{sss}
\end{equation}
is obtained from the real roots of the algebraic equation appearing 
in Eq. (\ref{alge}). A priori, depending upon the sign of $\Lambda$ and upon 
the relative weight of $\Lambda \epsilon$, $\rho_1$ can be either 
positive or negative. 

The solution derived in Eqs. (\ref{alge}) does not exhaust 
the possible critical points even in the case $x = y$. 
Consider, in fact, 
Eqs. (\ref{prim})--(\ref{ter}) in the case 
$ x = y$ but with $ z \neq 5\,x/2$. In this case Eq. (\ref{prim}) and 
Eq. (\ref{sec}) lead to the same algebraic equation
\begin{equation}
 2\,\L + 2\,\left( 5\,x^2 - 4\,x\,z + z^2 \right)  + 
\epsilon\,\left( -15\,x^4 + 48\,x^3\,z - 24\,x^2\,z^2 + 2\,z^4 \right)=0,
\label{prima}
\end{equation}
whereas Eq. (\ref{ter}) leads to
\begin{equation}
-2\,\L - 15\,x^2 + 10\,x\,z - 2\,z^2 + \epsilon\,\left( 45\,x^4 - 20\,x\,z^3 
+ 6\,z^4 \right)=0.
\label{seconda}
\end{equation}
Eqs. (\ref{prima})--(\ref{seconda}) cannot be further simplified 
and the best one can do is to solve them once the values of $\epsilon$ 
and $\Lambda$ are specified. 
If\footnote{The numerical roots reported in the present session are not just 
illustrative. They will be exploited in the numerical 
solutions discussed in the following Sessions.}  
$ \Lambda= -1$ and $\epsilon = 0.1$ 
the real roots of the system of Eqs. (\ref{prima})--(\ref{seconda}) 
 are given by:
\begin{eqnarray}
&&(x_1, \,z_1) = \pm(0.6974,\, 1.7436),\,\,\,\,\,
(x_2,\,z_2)= (\pm 1.3775,\mp 2.3429),
\label{root1}\\
&& (x_3,\,z_3)= (\pm 0.1338,\, \mp 
0.7176),\,\,\,(x_4,\,z_4) =(\pm 1.9645,\,\mp  0.5744).
\label{root2}
\end{eqnarray}
On top of the real roots reported in Eqs. (\ref{root1})--(\ref{root2}) there 
are also imaginary roots which are not relevant for the present discussion.
The first pair of roots in Eq. (\ref{root1})  can also be obtained 
from Eq. (\ref{alge}). Notice, also, that $(x_1,\,z_1)$ is 
the only root where the $z$ and $x$ can be simultaneously 
negative (when the minus sign is chosen). The other roots 
are such that $x$ and $z$ always have opposite signs.
The same discussion can be repeated for different values of $\L$ and 
$\epsilon$. For instance, if $\L = -2$ and $\epsilon= 0.1$ the real 
roots of Eqs. (\ref{prima})--(\ref{seconda}) are
\begin{eqnarray}
&&(x_1,\,z_1)=\pm(0.8857,\,2.2143),\,\,\,(x_2,z_2) = 
(\pm 1.3490 ,\, \mp 2.3077 ), 
\label{root3}\\
&&(x_3,\,z_3) =\pm ( 1.0275  ,\,  1.6067  ) , \,\,\, 
(x_4, z_4) = (\pm 0.2455 ,\, \mp 0.9111 ),
\label{root4}\\
&& (x_5,\,x_5) = \pm( 0.8563, 0.7023),\,\,\, 
(x_6, z_6) = (\pm 1.9285 ,\, \mp 0.5468 ).  
\label{root5}
\end{eqnarray}

If $x\neq y$ and, simultaneously, $ z \neq (4x + y)/2$ the system of 
Eqs. (\ref{prim})--(\ref{ter}) can still be solved.
From Eqs. (\ref{rel1}) and (\ref{rel2}) an {\em ansatz} for
the algebraic solution can be written as 
\begin{eqnarray}
&& z = \frac{ 3\epsilon x^2 -1}{6 \epsilon x},
\nonumber\\
&& y = \frac{ 1 - 9 \epsilon x^2 - 81 \epsilon^2 x^4 + 297 
\epsilon^3 x^6}{54 \epsilon^2 x^3 - 324 \epsilon^3 x^5}.
\label{ans1}
\end{eqnarray}
By inserting this {\em ansatz} back into Eqs. (\ref{prim})--(\ref{ter}) 
the same equation for $x$ is obtained from all three 
equations of the system, namely, 
\begin{equation}
2\,\L + \frac{5}{12\,\epsilon} + \frac{1}{648\,\epsilon^3\,x^4} + 
  \frac{1}{27\,\epsilon^2\,x^2} + \frac{7\,x^2}{3} + 
  \frac{25\,\epsilon\,x^4}{8} =0.
\end{equation}
Thus, the critical points of the system of Eqs. (\ref{prim})--(\ref{ter}) 
[in the case $ x\neq y $, $ x\neq 0$ and $z \neq (4 x + y)/2$] are given by 
\begin{eqnarray}
&& z = \frac{ 3\epsilon x^2 -1}{6 \epsilon x},\,\,\,\,
 y = \frac{ 1 - 9 \epsilon x^2 - 81 \epsilon^2 x^4 + 297 
\epsilon^3 x^6}{54 \epsilon^2 x^3 - 324 \epsilon^3 x^5},
\label{s2a}\\
&& P(x,\Lambda,\epsilon) =
\frac{1}{648\,\epsilon^3} + \frac{x^2}{27\,\epsilon^2} 
+ \left( 2\,\L + \frac{5}{12\,\epsilon} \right) \,x^4 + \frac{7\,x^6}{3} 
+ \frac{25\,\epsilon\,x^8}{8} =0.
\label{s2b}
\end{eqnarray}
Eq. (\ref{s2b}) should 
be solved for real values of $x$. The critical points corresponding to 
negative (and real) roots of Eq. (\ref{s2b}) lead
 to exponentially decaying $\sigma(\rho)$. Once the real 
roots of Eqs. (\ref{s2b}) are obtained, Eqs. (\ref{s2a}) 
will give the wanted values of $y$ and $z$.
\begin{figure}
\begin{center}
\begin{tabular}{|c|c|}
      \hline
      \hbox{\epsfxsize = 7 cm  \epsffile{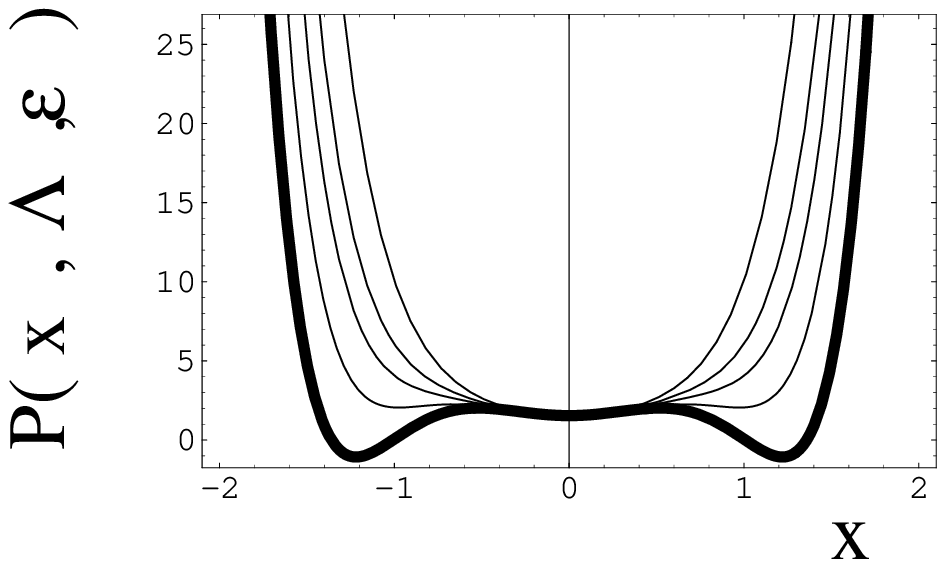}} &
      \hbox{\epsfxsize = 7 cm  \epsffile{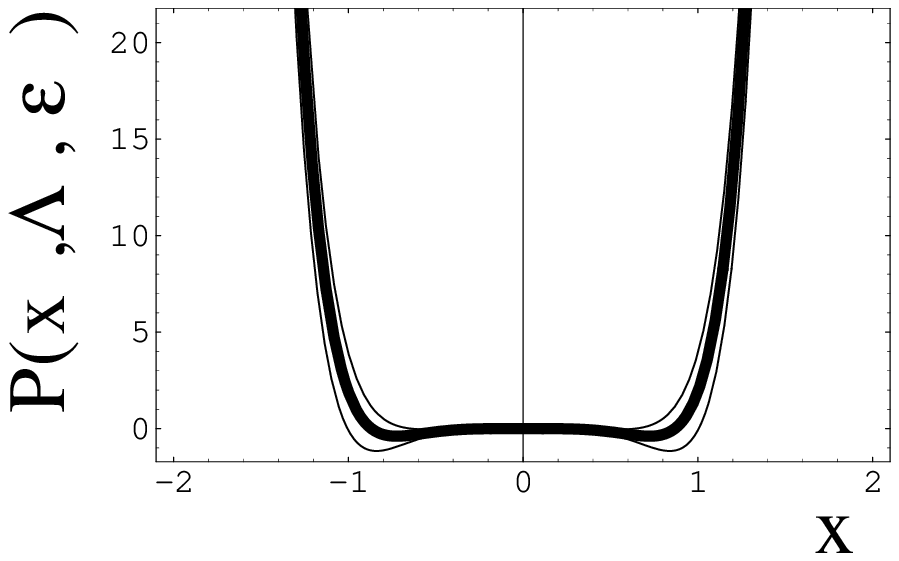}} \\
      \hline
\end{tabular}
\end{center}
\caption[a]{$P(x, \L,\epsilon)$ appearing in 
Eq. (\ref{s2b}) is plotted in the 
case of $\epsilon =0.1$ (left) and $\epsilon =1$ (right) and for 
different values of 
$\Lambda$. In the left plot the thick line corresponds to the case 
$\Lambda = -6$  where 
$P(x, \Lambda, \epsilon)$ has real roots. The other thin lines 
(from bottom to top) correspond, respectively, to $\Lambda = -5,-4,-3,-1$.
In the right plot the thick curve correspond to the case 
 $\Lambda =-2$ and the two thin lines correspond, from bottom to to top, 
to  $\L = -3$ and $\L = -1$. }
\label{fig1}
\end{figure}
Eq. (\ref{s2b}) does not always have real roots. In Fig. \ref{fig1} 
$P(x, \L,\epsilon)$ is reported as a function of $x$ and 
for different values of $\L$ and $\epsilon$. The specific 
values are only representative and will be used in specific numerical 
solutions later on. For instance, it can be argued from Fig. \ref{fig1}
that if $\epsilon = 0.1$ real roots of Eq. (\ref{s2b}) start appearing 
for $\L < -6$. 
 
\renewcommand{\theequation}{5.\arabic{equation}}
\setcounter{equation}{0}
\section{Stability of the critical points, 
singularities and numerical examples} 
If $x(\r)\equiv y(\r)$ [i.e. $\gamma(\r) = \gamma_0 \sigma(\rho)$],
Eqs. (\ref{ss1})--(\ref{ss2}) lead exactly to the same equation so that 
the only two independent equations obtained in this case are:
\begin{eqnarray}
&& a_x(\r)\, x' + a_z(\r)\, z' + a_0(\r) =0,
\nonumber\\
&& b_x(\r)\, x' + b_z(\r)\, z' + b_0(\r) =0.
\label{sysx=y}
\end{eqnarray}
where 
\begin{eqnarray}
&& a_x(\r) = 8 [ 1 - 3 \epsilon x ( x - 2 z) ],\,\,\,\,a_z(\r) = - 4 ( 1 - 
6 \epsilon x^2),
\nonumber\\
&& b_x(\r) = -10 ( 1 - 6 \epsilon x^2),\,\,\,\,\,\, b_{z}(\r) = 
4 ( 1 - 6 \epsilon z^2),
\end{eqnarray}
and 
\begin{eqnarray}
&& a_0(\r) = [2 \L + 10 x^2 - 8 x z + 2 z^2 +
\epsilon( -15 x^4 + 48 x^3 z - 24 x^2 z^2 + 2 z^4)],
\nonumber\\
&& b_0(\r) =[ - 2 \L - 15 x^2 + 10 x z - 2 z^2 + 
\epsilon ( 45 x^4 - 20 x z^3 + 6 z^4)].
\end{eqnarray}
Notice that the 
If $\Lambda =0$, 
from Eq. (\ref{alge})  $\rho_1$ becomes purely imaginary.
If $\L <0$ and if  $\r_1$ has to be real there are two relevant critical 
points, namely
\begin{equation}
\r_1 = \pm \sqrt{ \frac{5}{8 \L_{-}}} 
\sqrt{1 + \sqrt{1 + \frac{212}{5} \L_{-} \epsilon}},
\end{equation}
where, for convenience, $\Lambda = - \L_{-}$ (with $\L_{-} >0$).
In order to have a warped compactification scheme, $\r_1 <0$ and, therefore,
only one critical point satisfies this requirement.

If, as a warm-up, $\epsilon \rightarrow 0$ 
a critical point of the system is indeed given by Eq. (\ref{trecrit}). 
The tree-level critical point is not stable. Indeed, 
in the case $\epsilon =0$ Eqs. (\ref{sysx=y}) can be written 
as 
\begin{equation}
x' = f_x( x, z) ,\,\,\,\,\, z'= f_z(x, z),
\end{equation}
where 
\begin{eqnarray}
&& f_x(x,z) = x z - \frac{5}{2} x^2, 
\nonumber\\
&& f_z(x, z) = \frac{\Lambda}{2} - \frac{5}{2} x^2 + \frac{z^2}{2}.
\end{eqnarray}
The partial derivatives of $f_x(x,z)$ and $f_z(x,z)$ (i.e. $\partial_x f_x$,
$\partial_z f_x$ and $\partial_x f_z$, $\partial_z f_z$) define a matrix 
whose entries should be evaluated in the critical points, corresponding, in the
present case, to $ z = (5/2) x$, $x = - \sqrt{ 4 \L_{-}/5}$. The eigenvalues 
of this matrix are simply given by $\delta_{1, \,2} = \mp\sqrt{5 \L_{-}/4}$.
Therefore, the tree-level critical point is an unstable node since the 
eigenvalues are both real and with opposite 
signs (i.e.  $\delta_1 \delta_2 <0$). 

Suppose now to turn on the first $\alpha'$ correction. Then 
$\epsilon \neq 0$.  
In this case the tree-level discussion can be repeated. If $\epsilon \neq 0$
the dynamical system will now become [always in the case $x(\r) = y(\r)$]
\begin{equation}
x' = g_x( x, z) ,\,\,\,\,\, z'= g_z(x, z),
\end{equation}
where now, from eq. (\ref{sysx=y})
\begin{eqnarray}
&& g_x(x,z) = \frac{a_z b_0 - a_0 b_z}{b_x a_z - a_xb_z},
\nonumber\\
&& g_z(x,z) = \frac{a_x b_0 - a_0 b_x}{b_x a_z - a_xb_z}.
\end{eqnarray}
In analogy with the previous case the matrix of the derivatives can be 
constructed.
The two eigenvalues of such a matrix are complex conjugate
numbers of the form 
\begin{equation}
W_{1,\,2} = \sqrt{\L_{-}}[\mu(k) \pm i ~\nu(k)].
\label{eigenv}
\end{equation}
where $k = \L_{-} \epsilon$ and the subscripts $1$ and $2$ correspond, 
respectively, to the plus and minus signs.
The expressions are quite cumbersome so that they are 
reported in the Appendix C. In order to determine the stability 
of the system the sign of $\mu(k)$ is crucial. It turns out, from a 
numerical analysis of Eqs. (\ref{w1})--(\ref{w2}), that
\begin{eqnarray} 
\mu(k) >0,\,\,\,\,{\rm for}\,\,\,k = \L_{-}\epsilon > 0.65,
\nonumber\\
 \mu(k) <0,\,\,\,\,{\rm for}\,\,\,k = \L_{-}\epsilon < 0.65.
\label{eigenv2}
\end{eqnarray}
Thus,  for $\L_{-}\epsilon > 0.65$  there is an unstable spiral point
whereas for $\L_{-} \epsilon <0.65$ there is a stable spiral point \cite{dyn}.

In closing this session an interesting solution 
can be mentioned. In the case $\L = 0$ and $z=0$ (constant dilaton case)
the relevant equations can be written as :
\begin{eqnarray}
&&6 [ 1 - \epsilon \, x( x + 2 y)] x' + 2 [ 1 - 3 \epsilon x^2] y' 
+ 2 \L + 6 x^2 
\nonumber\\
&&+ y ( y + 3 x) - 3 \epsilon x^2 ( y^2 + 3 x y + x^2) =0,
\label{on}\\
&&8 ( 1 - 3 \epsilon x^2 ) x' + 2 \L + 10 x^2 - 15 \epsilon x^4=0.
\label{tw}
\end{eqnarray}
The critical point of the system can be obtained also in this case 
and it corresponds to $ x_{c} = y_{c} = -\sqrt{ 2/( 3 \,\epsilon)}$.
The system can then be written in 
 its canonical form, namely, $ x'= F_{x}(x,y)$ and $ y'= F_{y}(x,y)$. 
The eigenvalues of the matrix of the derivatives evaluated in $ (x_{c},\,y_c)$
are, respectively, $\delta_1 = 5/\sqrt{ 6\epsilon}$ and $ \delta_2 = 
5 ( 3 \epsilon -1)/\sqrt{ 6\epsilon}$. Notice that $\delta_1 \delta_2 = 
[25(3 \epsilon -1)]/(6\epsilon)$. This means that the critical point 
is a saddle point for $\epsilon < 1/3$ and an unstable node for $\epsilon >1/3$
\cite{dyn}. This solution, even though interesting, is only illustrative : 
the presence of the inverse coupling in the critical point clearly points 
towards non-perturbative effects which should be properly 
discussed by adding higher orders in $\alpha'$.

\subsection{Numerical examples}

A Kasner-like branch can be analytically connected to 
a constant curvature solution where $ x_0 <0$ and the dilaton 
is linearly decreasing ($ z_0 < 0$).
Suppose
that $\L = -1$ and $\epsilon =0.1$. Then, we can see that there exist 
a solution connecting the Kasner-like regime to a 
critical point given by Eqs. (\ref{alge}) and (\ref{root1}). 
\begin{figure}
\begin{center}
\begin{tabular}{|c|c|}
      \hline
      \hbox{\epsfxsize = 7 cm  \epsffile{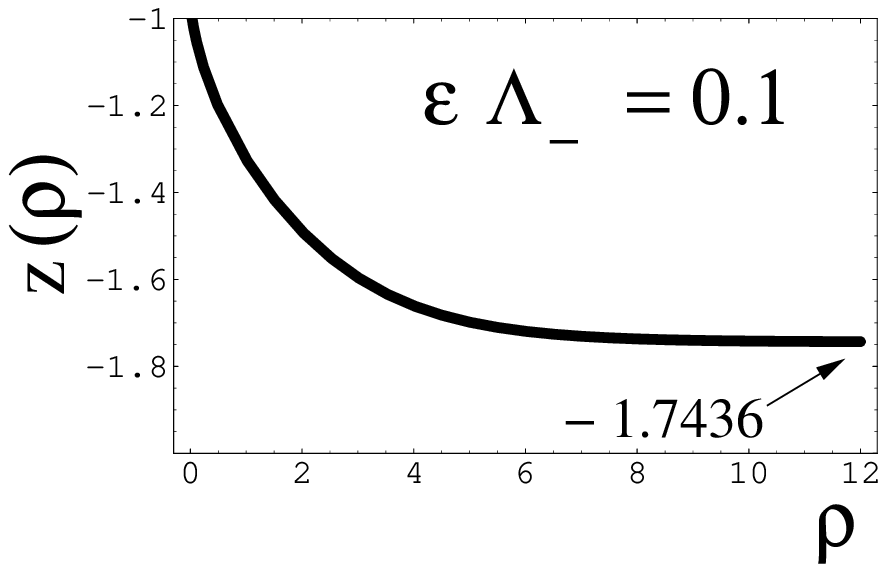}} &
      \hbox{\epsfxsize = 7 cm  \epsffile{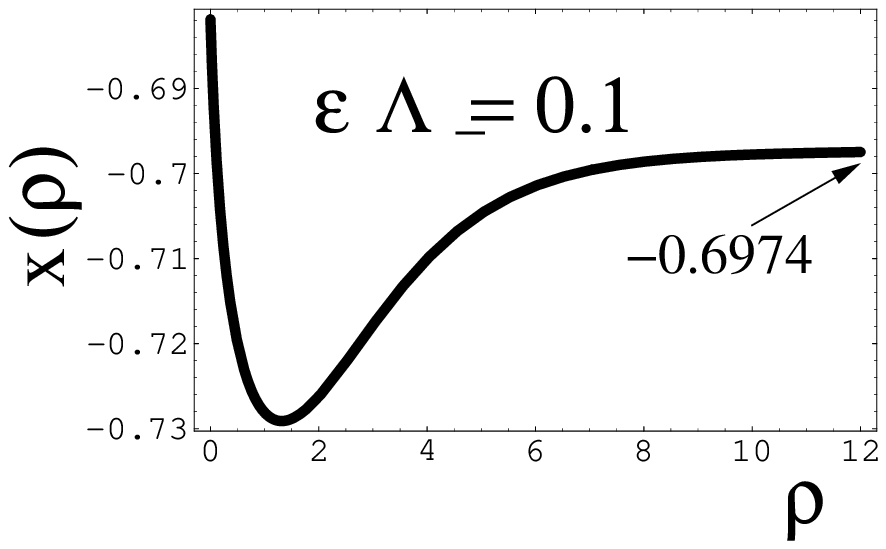}} \\
      \hline
\end{tabular}
\end{center}
\caption[a]{A numerical solution of the system given in Eq. (\ref{sysx=y}) 
is reported 
in the case $\L = -1$ and $\epsilon = 0.1$. The solution 
interpolates between the Kasner-like regime (for $\r \rightarrow 0$)
and the critical point given by the solution 
of Eq. (\ref{alge}) which is explicitly given (for the chosen values of 
$\L$ and $\epsilon$)  by the first pair of roots appearing 
in Eq. (\ref{root1}).}
\label{fig2}
\end{figure}
The numerical result is reported in Fig. \ref{fig2}. For large 
$\rho$ the solution matches exactly the critical point
which can be obtained (with $\L= -1$ and $\epsilon = 0.1$) from 
Eq. (\ref{alge}) and which are reported in the first pair of roots 
in Eq. (\ref{root1}).

If the initial conditions are not given in the Kasner-like 
regime but in the vicinity of a critical point there are 
various possibilities. It can happen that the system 
is driven (smoothly) towards another critical point or it can happen 
that a singularity is reached.

\begin{figure}
\begin{center}
\begin{tabular}{|c|c|}
      \hline
      \hbox{\epsfxsize = 7 cm  \epsffile{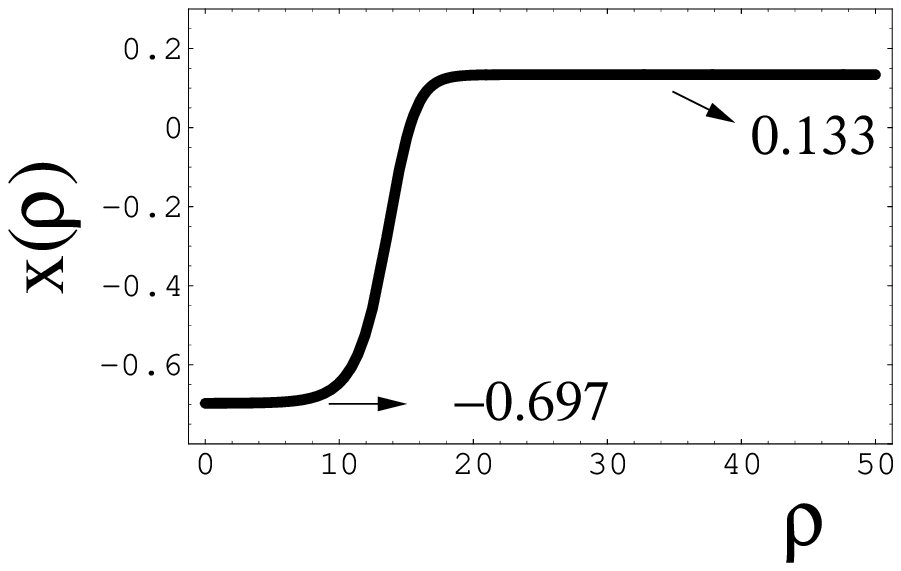}} &
      \hbox{\epsfxsize = 7 cm  \epsffile{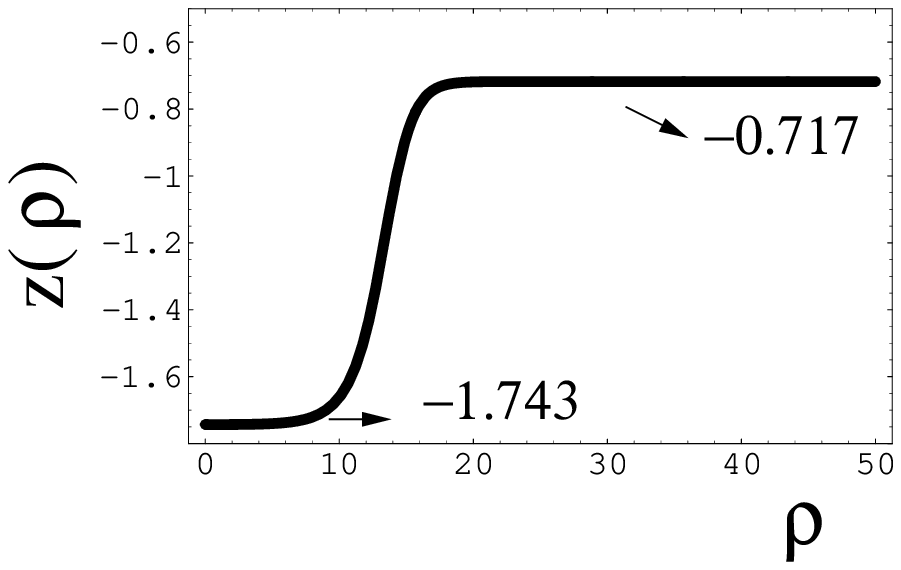}} \\
      \hline
\end{tabular}
\end{center}
\caption[a]{The numerical integration of Eqs. (\ref{sysx=y}) 
is reported in the case 
$\L= -1$ and $\epsilon = 0.1$. The critical points in this specific case 
were given explicitly in Eqs. (\ref{root1})--(\ref{root2}). Initial conditions
are given in one of them, namely $(x_1,\,z_1)$. }
\label{fig3}
\end{figure}

Suppose that the  initial conditions for the numerical integration 
of the system of Eq. (\ref{sysx=y}) are given in in $ \rho \sim 0$ requiring 
that $ x(0) <0 $ and $ z(0)<0$ according to the solution of Eqs. (\ref{alge}).
In this case, the system seats exactly on a critical 
point whose properties have been previously analyzed.
In the case  $\Lambda = -1$ and $\epsilon = 0.1$ the initial conditions are 
fixed for
$x(0)\equiv x_1$ and $z(0) \equiv z_1$ where 
$(x_1,\,z_1)$ are given by Eq. (\ref{root1}). In this case 
 the system evolves towards a different critical point, 
namely it happens 
that $x(\infty) \rightarrow x_3 $ and $z(\infty) \rightarrow z_3 $ where 
$(x_3,\,z_3)$ are given by Eq. (\ref{root2}). This 
conclusion can be obtained by direct numerical 
integration. In Fig. \ref{fig3} the numerical integration is reported for the 
mentioned initial conditions. It can be seen that 
$x(50) = 0.133$ and $z(50) =-0.7176$ which, indeed, coincide with 
$x_3$ and $z_3$ within the numerical accuracy 
of the present analysis.
The singularity structure of the curvature invariants 
can be appreciated by analyzing the behavior of
\begin{eqnarray}
&& I_1(\r) =R_{\mu\nu\alpha\beta} R^{\mu\nu\alpha\beta}
,\,\,\,\,I_2(\r) = C_{\mu\nu\alpha\beta} C^{\mu\nu\alpha\beta}
\nonumber\\
&& I_3(\r) = R,\,\,\,\ I_4(\r) = R_{\mu\nu}R^{\mu\nu}
\end{eqnarray}
whose 
general form is reported, for the metric (\ref{line}), in Appendix C.
Notice that if $x = y$, as in the example of Fig. \ref{fig3}, the Weyl
invariant vanishes. 
\begin{figure}
\begin{center}
\begin{tabular}{|c|c|}
      \hline
      \hbox{\epsfxsize = 7 cm  \epsffile{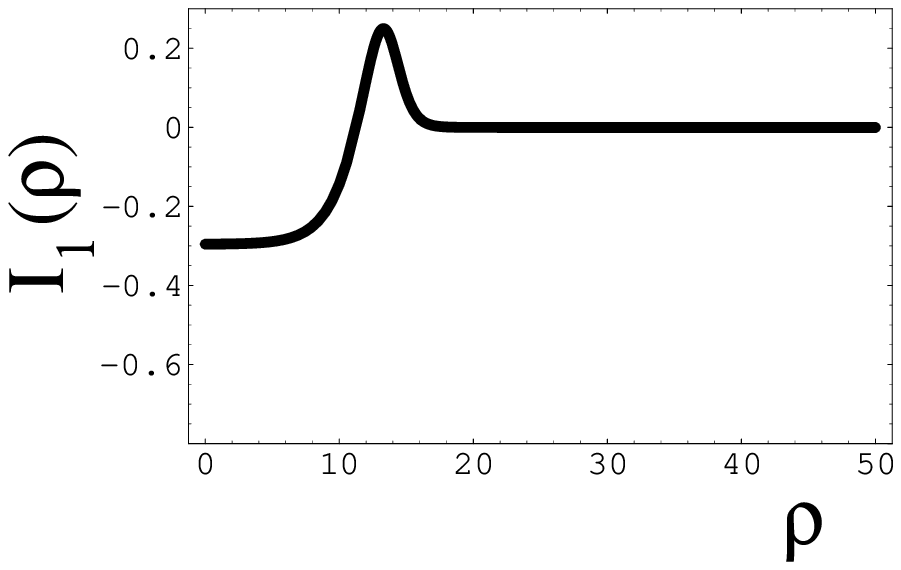}} &
      \hbox{\epsfxsize = 7 cm  \epsffile{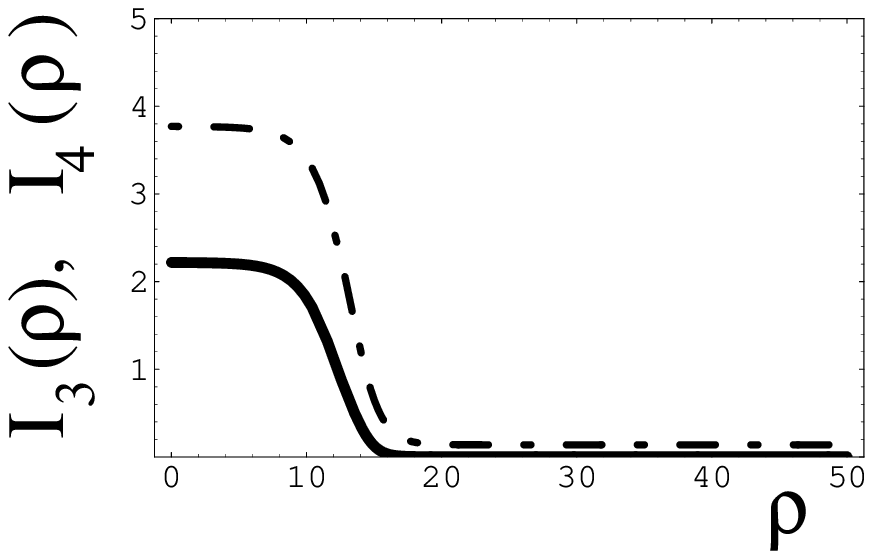}} \\
      \hline
\end{tabular}
\end{center}
\caption[a]{In the case of $\L = -1$ and $\epsilon = 0.1$ 
the curvature invariants corresponding to the solution reported in Fig. 
\ref{fig3} are reported.
The  Riemann invariant appears in the left plot with full thick line. 
At the right the Ricci invariant is reported  with dot-dashed line. 
The  scalar curvature is reported (always in the right plot)
 with the full (thick) line.}
\label{fig3a}
\end{figure}

Suppose now that $\Lambda = -2 $ and $\epsilon = 0.1$. The critical points 
are given by Eqs. (\ref{root3})--(\ref{root5}). The initial conditions imposed 
in $ \r = 0$ will be such that $ x(0) <0$ and $z(0) <0$ as in the previous 
numerical example. Requiring, from Eq. (\ref{root3}), $x(0) = x_1$ and 
$z(0) = z_1$ the numerical integration leads to a divergence for large $\rho$. 
In this specific case, the singularity occurs around 
$\rho \sim 6$.
\begin{figure}
\begin{center}
\begin{tabular}{|c|c|}
      \hline
      \hbox{\epsfxsize = 7 cm  \epsffile{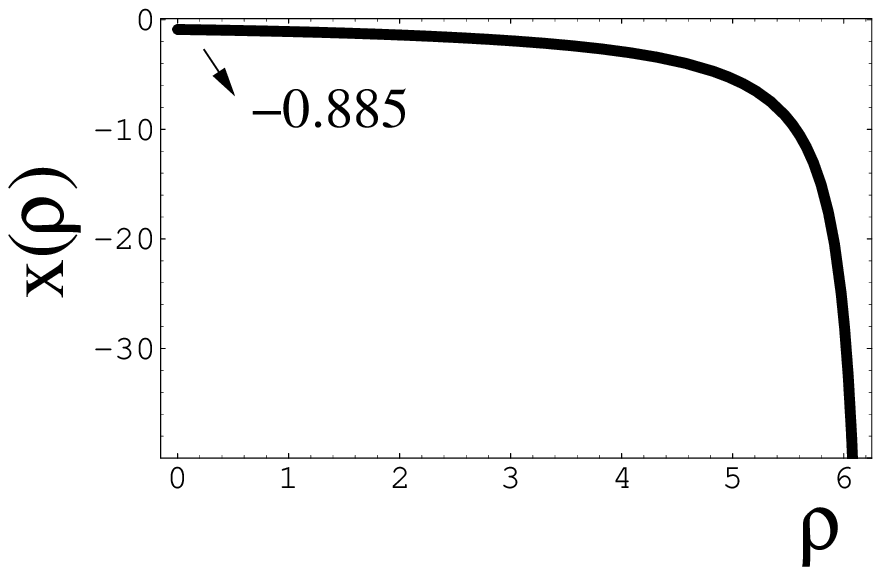}} &
      \hbox{\epsfxsize = 7 cm  \epsffile{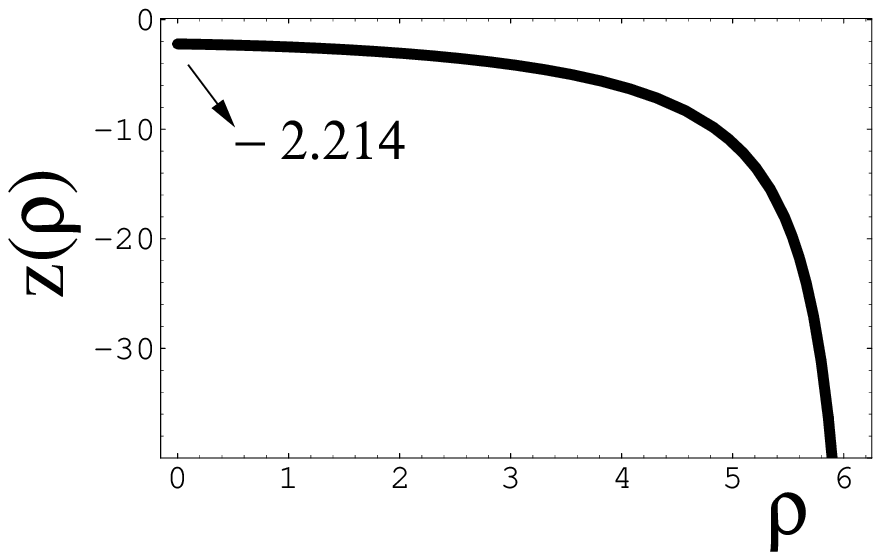}} \\
      \hline
\end{tabular}
\end{center}
\caption[a]{In the case $\L = -2$ and $\epsilon = 0.1$ the numerical
 integration
of the system (\ref{sysx=y}) is reported. The initial conditions are given 
in a critical point of the system , namely the point $(x_1,\,z_1)$ reported in 
Eq. (\ref{root3}).}
\label{fig4}
\end{figure}
Again, the proof of this statement can be given by direct 
numerical integration whose results are 
reported in Fig. \ref{fig4}.
\begin{figure}
\begin{center}
\begin{tabular}{|c|c|}
      \hline
      \hbox{\epsfxsize = 7 cm  \epsffile{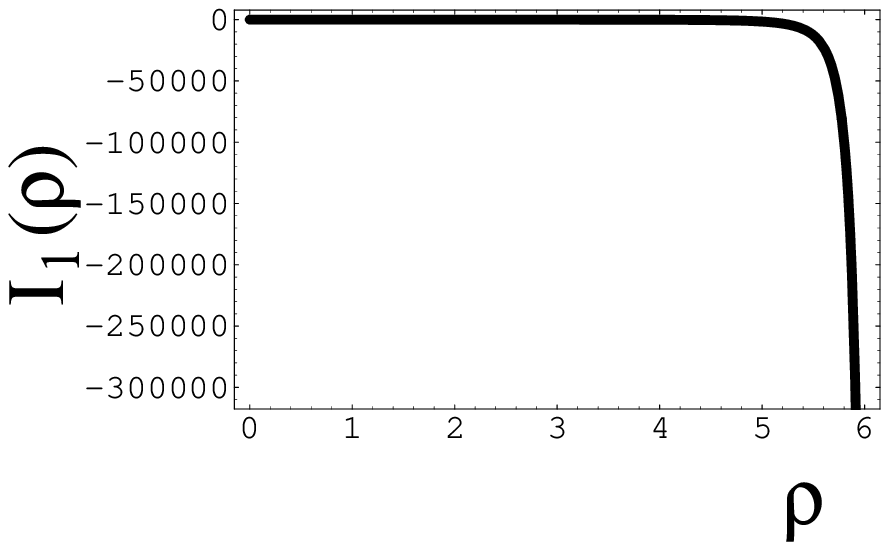}} &
      \hbox{\epsfxsize = 7 cm  \epsffile{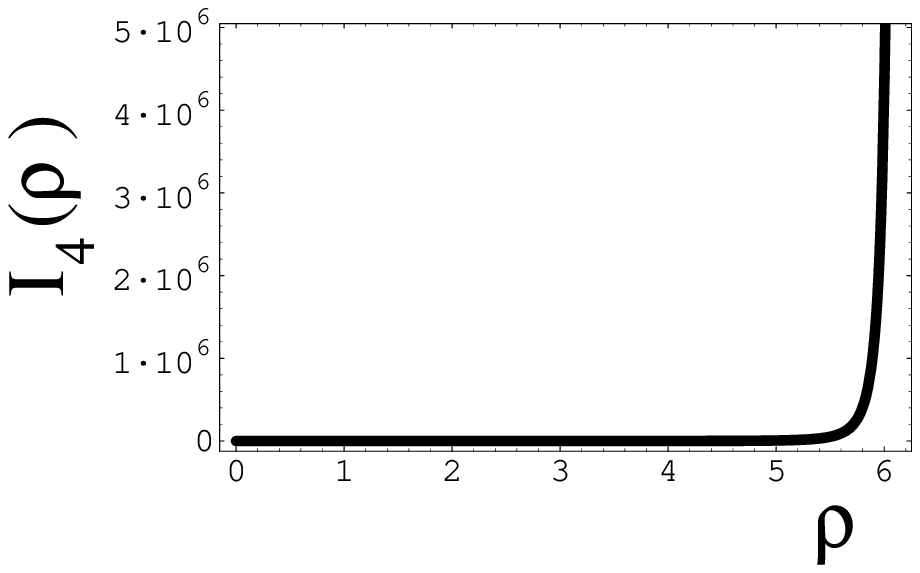}} \\
      \hline
\end{tabular}
\end{center}
\caption[a]{The Riemann (left) and Ricci (right) invriants are 
reported for the numerical solution obtained in Fig. \ref{fig4}.}
\label{fig4a}
\end{figure}
Numerical examples concerning the case 
$x \neq y$ will now be given. The specific values of $\epsilon$ and $\Lambda$ 
are just meant to illustrate some interesting features of the solutions.
As a general remark we could say that in the case $x \neq y$ 
singularities seem to be more likely. This statement is
only based on a scan of the numerical solutions for different 
values of $\L$ and $\epsilon$.
 
Suppose that $\epsilon =0.1$. Thus, from the examples 
of Fig. (\ref{fig1}), $\L$ needs to be sufficiently negative in order 
to produce real (negative) roots in Eq. (\ref{s2b}).
Suppose, for instance that $\Lambda = -10$. Then, 
from Eqs. (\ref{s2a})--(\ref{s2b}), the 
critical points will be, 
\begin{eqnarray}
&&(x_1,\,y_1,\,z_1)=(-2.0258,\,0.6386,\,-0.1902),\,\,\,
\nonumber\\
&&(x_2,\,y_2,\,z_2) =
(-0.6889,\,-3.3432,\,2.0748 ).
\label{root6}
\end{eqnarray}
\begin{figure}
\begin{center}
\begin{tabular}{|c|c|}
      \hline
      \hbox{\epsfxsize = 7 cm  \epsffile{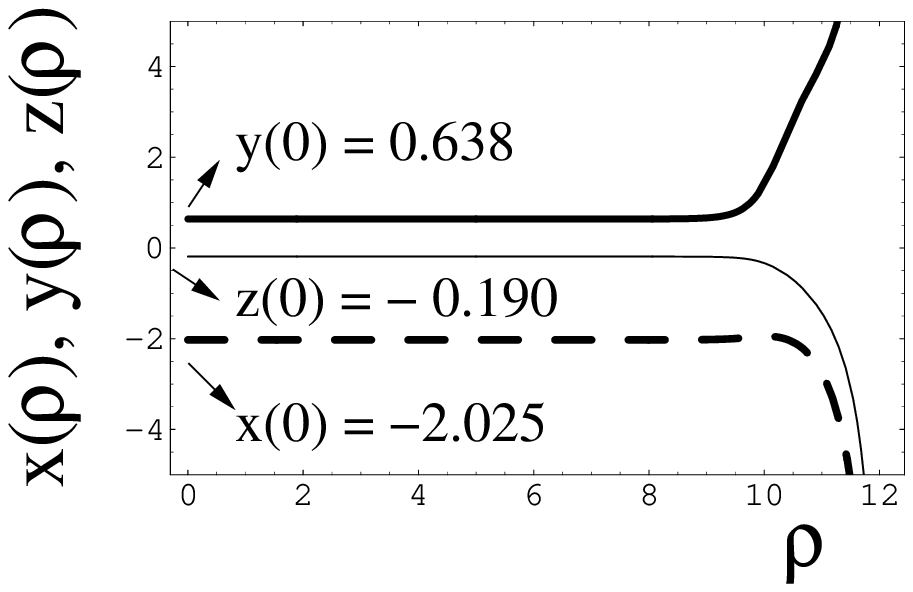}} &
      \hbox{\epsfxsize = 7 cm  \epsffile{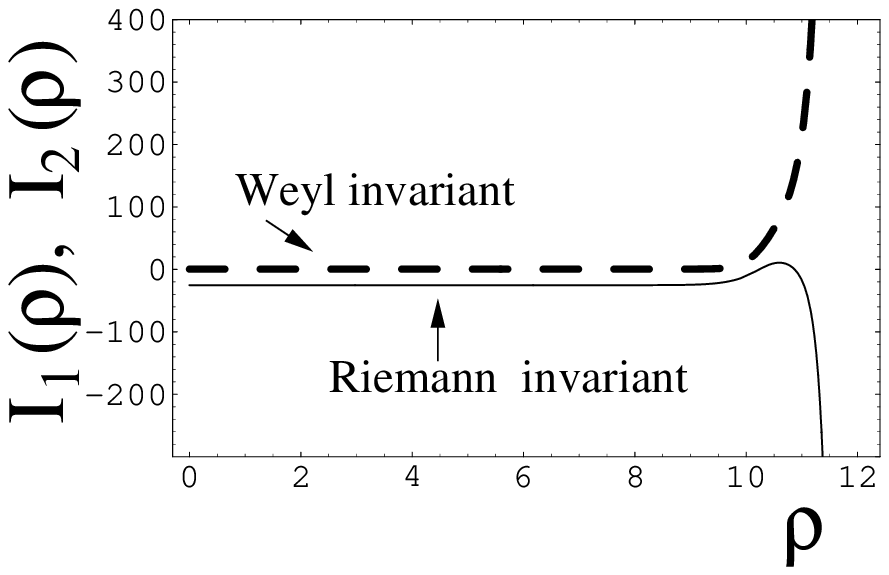}} \\
      \hline
\end{tabular}
\end{center}
\caption[a]{In The case of $\L = -10$ and $\epsilon = 0.1$ 
the numerical integration of the system (\ref{sys1}) is reported in the 
left plot. The evolution of $x$ (dashed line), $y$ (full thick line) and 
$z$ (thin line) are illustrated when the initial conditions are given in 
$(x_1,y_1,z_1)$ of Eq. (\ref{root5}). In the plot at the right the Weyl 
(dashed line) 
and the Riemann (full line) are reported.}
\label{fig5}
\end{figure}
If the initial conditions for the warp factors and for the dilaton are 
set in $(x_1,\,y_1,\,z_1)$ of Eq. (\ref{root5})
 then the results of the numerical integration 
are reported in Fig. \ref{fig5}. A physical singularity 
is developed at a finite value of $\rho$ where both the Weyl and the Riemann 
invariant blow up. 
The same kind of solutions can be obtained if the initial conditions are set 
not in $(x_2,\,y_2,\,z_2)$ of Eq. (\ref{root6}). Also in this second case 
a singularity is developed at finite $\rho$.

In order to describe the qualitative behavior of the system 
the value of $\epsilon$ can be changed. From Fig. \ref{fig1} (left plot) 
$\L= -2$ can be selected and from Eqs. (\ref{s2a})--(\ref{s2b}) the critical
 points  are found to be 
\begin{eqnarray}
&&(x_1,\,y_1,\,z_1) = (-0.8679,\,0.6052,\,-0.2419),
\nonumber\\
&&(x_2,\,y_2,\,z_2) = (-0.1638,\,-3.5390,\,0.9350).
\label{root7}
\end{eqnarray}
\begin{figure}
\begin{center}
\begin{tabular}{|c|c|}
      \hline
      \hbox{\epsfxsize = 7 cm  \epsffile{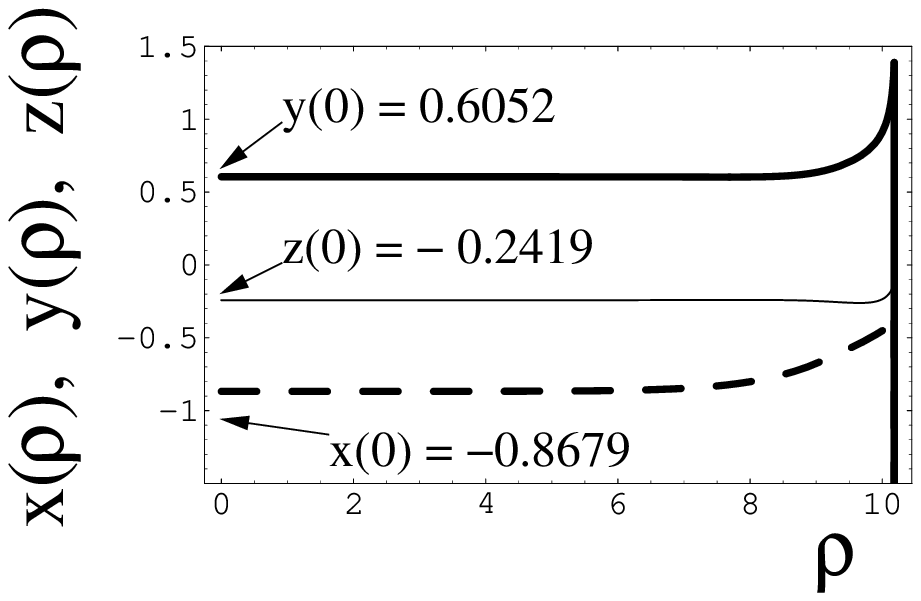}} &
      \hbox{\epsfxsize = 7 cm  \epsffile{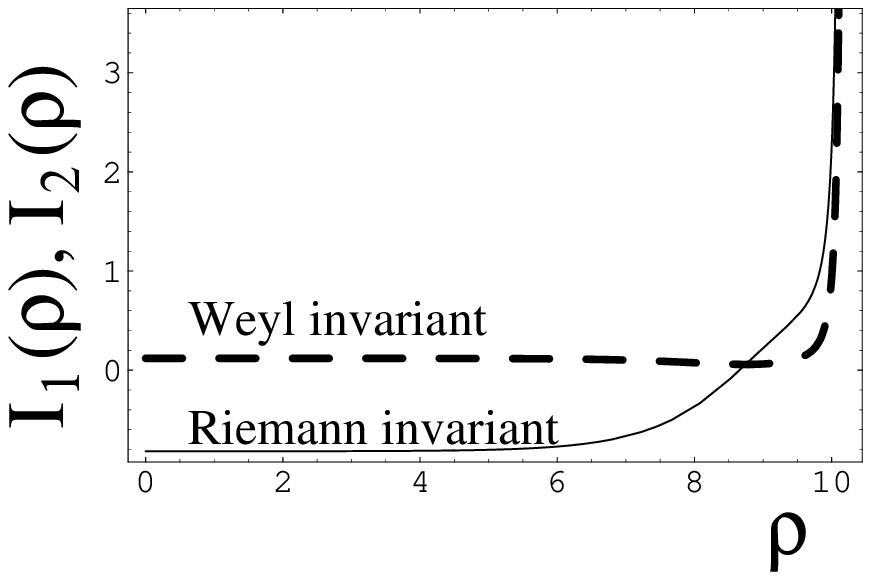}} \\
      \hline
\end{tabular}
\end{center}
\caption[a]{In the left plot the numerical integration 
of eqs. (\ref{sys1}) is illustrated for $\L= -2$ and $\epsilon = 0.1$. 
The notations
are exactly the same as in Fig. \ref{fig5}. The initial conditions 
are given in the point $(x_1,y_1,z_1)$ reported in Eq. (\ref{root7}). }
\label{fig6}
\end{figure}
In this case the results of the numerical integration are 
reported in Fig. \ref{fig6}. Also in this example 
a physical singularity is developed at finite value of $\r$.

In the present Section we examined systematically 
 the critical points of the dynamical
system arising from the 
low energy $\beta$ functions truncated to first order in the 
string tension corrections. For practical purposes two cases 
have been distinguished namely the one where 
the warp factors are proportional and the one where the warp 
factors are both exponentially decreasing but not proportional to each other.
From the analysis of both cases we can conclude that there are no (stable) critical 
points of the $\beta$-functions where the dilaton is frozen. This 
means that stable (or unstable) critical points with 
decreasing dilaton field cannot be analytically connected with 
solutions where the dilaton coupling is frozen to a constant value.
However, a large class of critical points leads to warped compactifications in the sense 
that the warp factors are exponentially decreasing and the dilaton coupling 
is also exponentially decreasing (without getting fixed to a constant value).

As pointed out in the introduction, the present investigation suffers from 
various limitations. That is why the present results 
can only be considered as illustrative. However, it is not unreasonable 
that some of the nasty features of the present results might be 
changed if thick branes are included. In this case  
some of the unstable solutions might get stabilized since a new scale 
(the thickness of the brane) arises in the problem. For instance 
solutions with exponentially decreasing warp factors and decreasing 
dilaton coupling might turn into warped solutions with constant 
dilaton. On the contrary warped solutions with increasing 
dilaton coupling (corresponding to another class of fixed points 
emerging from the present analysis) might be driven towards 
naked singularities with Kasner-like features. 

\renewcommand{\theequation}{6.\arabic{equation}}
\setcounter{equation}{0}
\section{Concluding Remarks} 

The effects of the dilaton and of higher order curvature terms 
are usually neglected in the context of six dimensional warped 
compactification. This is certainly a consistent assumption
which can be, however, relaxed. 
In the present investigation six-dimensional 
warped compactification has been analyzed in the context 
of the gravity theory inspired by the low energy string effective action.
Tree-level solutions have been derived. First order $\alpha'$ 
corrections have been also included precisely in the form 
induced by string amplitudes.

The tree-level solutions
 exhibit a Kasner-like branch whose 
generic property is the occurrence of curvature singularities.
In this case the warp factors are powers of the bulk 
radius $\r$. The powers satisfy specific sum rules leading ultimately 
to divergences in the curvature invariants. 
The
features of the tree-level solutions (as a function of the bulk radius) 
are the static analog of the time-dependent case 
which has been widely exploited in cosmological considerations.
Kasner-like solutions can be found both for vanishing and non 
vanishing bulk cosmological constant. This means that, at tree-level, 
exponentially decreasing warp factors cannot be generically 
obtained. Moreover, since the Kasner-like behaviour leads 
to curvature singularities, there will be a regime where 
the solutions hit the string curvature scale where 
higher order string tension corrections cannot be neglected.

The inclusion of the first order $\alpha'$ correction produces 
computable modifications in the $\beta$-functions which can be 
viewed, for the present purpose, as a nonlinear dynamical system.
Solutions with exponentially decreasing warp factors 
become then possible: they correspond to critical points of 
the dynamical system. 
Defining ${\cal H}= \partial_{\rho} \ln{\sigma} $ and ${\cal F} = 
\partial_{\rho} \ln{\gamma}$ interesting critical points can be obtained 
in the case of constant (and negative) ${\cal H}$ and ${\cal F}$. These 
critical points correspond to linear (decreasing) dilaton solutions.
  
The obtained critical points are not always stable. It can happen 
that a given critical point correspond to  stable/unstable node 
or to stable/unstable spiral points. In the case ${\cal H} = {\cal F}$ 
the Kasner branch of the solution (for $\r \sim 0$)  
can be analytically connected to a critical point.
There are also solutions where two critical points 
can be analytically connected. In this case a given ${\cal H}_1 <0$ 
turns into ${\cal H}_{2} \neq {\cal H}_1$. Unfortunately in the numerical 
examples analyzed in the present investigation ${\cal H}_2  >0$.

In the case ${\cal H}\neq {\cal F}$ the situation is mathematically more 
complicated. By giving initial conditions of the system 
near a critical point singularities seem to be 
 developed. Numerical evidence of this behavior 
has been presented by studying the singularity properties of the curvature 
invariants. 

It is now the moment of comparing what we have with what we ought to have. The ideal 
situation would be to find stable fixed points of the low 
energy $\beta$ functions compatible with a constant dilaton field. 
In other words it would be nice to find solutions interpolating 
between a singularity and a critical point of constant dilaton or between 
an unstable critical point and a constant dilaton solution. Such a 
behavior would give rise to exponentially decreasing warp factors 
with constant dilaton coupling which is exactly the original assumption 
tacitly adopted in five and six-dimensional warped compactifications.
This situation is never realized in the set-up described 
in the present investigation but it cannot be excluded in principle in slightly 
different frameworks. 
It has been shown that exponentially decreasing 
warp factors are certainly compatible with the low energy string 
effective action, however, the obtained critical points do not have the 
wanted physical properties. These results should not be viewed
as conclusive and might only be due to the limitations of the present analysis.

We will now speculate about various possibilities which might 
cure the limitations of our approach allowing, hopefully, for 
stable critical points with frozen dilaton. As we discussed 
the first (obvious) limitation which should be relaxed 
concerns the dilaton dynamics. In the present paper only string tension 
corrections have been considered. In principle one can expect 
that corrections in the dilaton coupling should also be included for 
consistency since they become relevant roughly at the same 
scale at which string tension corrections are turned on. 
Following a complementary way of thinking one can 
also try to account for the effect of the dilaton potential 
which has been also (partially) neglected since we modeled it with 
a cosmological constant.  An encouraging remark, in this 
direction, would be  that already within our over-simplified set-up
decreasing dilaton solutions were compatible with exponentially decreasing 
warp factors. These solutions might be   
analytically connected with a critical point of constant dilaton once 
a dilaton potential is present. A less encouraging evidence is however that
 constant dilaton solutions are obtained (within the present analysis) only as 
$1/\alpha'$ effect. These solutions are clearly non-perturbative 
and it might be that only by summing up all the string tension corrections 
a definite answer can be obtained.

Another limitation of the present analysis is the absence of realistic (thick) 
brane sources. It is not unreasonable to expect that by adding a brane of 
finite thickness in the game some of the unstable critical 
points of the system will be stabilized at a constant value of the 
dilaton coupling fixed, presumably, by the tension of the brane. This 
possibility should be further scrutinized. Again an encouraging hint in this 
direction is that, already at the level of our analysis, it is possible to connect 
a critical point to Kasner-like solutions. 

Furthermore  in our analysis we completely 
neglected the possible presence of $p$-form fields whose effect 
should also be studied. To neglect the antisymmetric field strength is 
consistent if we look at the 
explicit form of the first string tension correction. 
However,  an antisymmetric tensor field
with large magnetic component might introduce qualitatively new solutions. 
In the present discussion the antisymmetric tensor field has been frozen. However, once 
an antisymmetric tensor is allowed on the same footing of the other light modes 
the structure of the string tension corrections changes radically: higher powers of the 
field strength will appear to first order in the string tension corrections.
We leave all these themes for forthcoming investigations.

Finally, some cosmological implications of our work should 
be mentioned. Once thick branes, antisymmetric tensor field and dilaton potetial will 
be taken into account in simplified models, time dependent solutions might be studied. 
In this case the problem will certainly be more complicated but not totally 
hopeless if stable critical points of the truncated $\beta$ functions 
will be found in the time-independent case. We also leave these 
topics for future investigations. 
  
\section*{Acknowledgments}
The author wishes to thank M. E. Shaposhnikov for interesting discussions.

\newpage

\begin{appendix}
\renewcommand{\theequation}{A.\arabic{equation}}
\setcounter{equation}{0}
\section{Curvature Invariants}
In order to scrutinize the singularity properties of 
the six-dimensional metric discussed in the 
present investigation the quadratic curvature 
invariants should be properly discussed. 
The curvature invariants computed from Eq. (\ref{line}) 
can be written as 
\begin{eqnarray}
&& I_1(\r) = R_{\mu\nu\alpha\beta} R^{\mu\nu\alpha\beta} =
\frac{1}{4}[ \F^4 - 6 \H^4 + 4 \F^2 \F' + 4 \F'^2 
+ 16 \H^2 \H' + 16 \H'^2]
\label{riemannsq}\\
&& I_2(\r) =C_{\mu\nu\alpha\beta} C^{\mu\nu\alpha\beta} = \frac{3}{20} 
[ 2 (\H' - \F') + \F( \H - \F)]^2
\label{weylsq}\\
&&I_3(\r)= R = \frac{1}{2}[ \frac{3}{2}
\F^2 + 4 \H\F +10 \H^2 + 2\F' + 8 \H']
\label{scalarc}\\
&& I_4(\r) =R_{\mu\nu} R^{\mu\nu} = \frac{1}{8} [ \F^4 + 4 \H \F^3 
+ 14 \H^2 \F^2 + 16 \H^3 \F + 40 \H^4 + 4 \F^2 \F' 
\nonumber\\
&&+ 
8 \H \F \F' + 8 \H^2 \F' + 4 \F'^2 + 8 \F^2 \H' 
+ 8 \H \F \H' + 64 \H^2 \H' + 16 \H' \F' + 40 \H'^2]
\label{riccisq}
\end{eqnarray}
The Weyl invariant vanishes in the case where 
$\sigma(\r)$ and $\gamma(\r)$ are proportional, namely in 
the case where $\H \equiv \F $. The other invariants 
are do not vanish in the limit $\H\rightarrow \F $ .

The curvature invariant are singular in the case of the
tree-level solutions derived in Eqs. (\ref{tlev}).
For instance the Riemann and Weyl invariants can be written, for 
the solutions of Eqs. (\ref{tlev}), as:
\begin{eqnarray}
&& I_1(\r)= \frac{\L^2}{16}\,
[ 4\,\a^2 - 12\,\a^4 + \b^2 + 2\,\b^4 - 
    4\,\left( 4\,\a^3 + \b^3 \right) \,
     \cosh{[{\sqrt{-\L}}\,(\r-\r_0)]} 
\nonumber\\
&&+ 
    \left( 4\,\a^2 + \b^2 \right) \,\cosh{[2\,{\sqrt{-\L}}\,(\r-\r_0)]}]
   \,{\sinh{[{\sqrt{-L}}\,(\r-\r_0)]}}^{-4},
\nonumber\\
&& I_2(\r) = 
\frac{3}{40}\left( \a - \b \right)^2\,\L^2\,
  {\left( \b - \cosh{[{\sqrt{-\L}}\,(\r-r_0)]} \right) }^2\,
   {\sinh{[{\sqrt{-\L}}\,(\r -\r_0)]}}^{-4}.\, 
\end{eqnarray}
This shows that the tree-level (Kasner-like) solutions are 
indeed singular for $\rho\rightarrow \r_0$. Notice, again, that 
if $\alpha = \beta$ the solution still exists. In this 
case the Weyl invariant vanishes identically but the Riemann
invariant (and the other invariants) are still singular.

In order to detect singularities 
in a  given solution the behavior of the curvature invariants 
has been scrutinized as a function of the bulk radius. This is 
a consistent procedure. In order to fully analyze the singularity
properties of the solution
we should also investigate the behavior of non-space-like geodesics 
(i.e. either time-like or null). This check is pleonastic 
if the curvature invariants diverge. However, the regularity 
of the curvature invariants is a necessary but not sufficient condition 
in order to assess the regularity of a manifold. In principle, in order to 
claim that a given geometry is singularity-free we should chack that 
the curvature invariants are regular and that 
non-space-like geodesics are complete (i.e. they can be extended to any 
value of the affine parameter). These are the usual general relativity
criteria for the analysis of physical singularities of a given space-time. 
Using Einstein equations together with topological considerations these 
criteria can be rephrased in terms of energy conditions involving the various 
components of the energy momentum tensor. Hawking-Penrose theorems 
are based on this type of considerations \cite{HP}.

When singularities are analyzed in the context of the low-energy string effective action 
there is a further ambiguity related to the choice of the physical frame where 
the calculation is performed.
In this paper the string frame parameterization of the action has been used.  
Other frames can be anyway defined. A very useful frame is the Einstein frame where 
the dilaton and the graviton are decoupled (at tree-level). Defining 
as $g_{\mu\nu}$ a $D= 4 + n$ dimensional metric in the string frame and 
as $G_{\mu\nu}$ as the Einstein frame metric (in the same number of 
dimensions) we have that they are related through a conformal rescaling 
involving the string frame dilaton $\phi$, namely:
\begin{equation}
G_{\mu\nu} = e^{-\frac{2 \phi}{n + 2}} g_{\mu\nu}.
\label{transf2}
\end{equation}
The (canonically normalized) 
Einstein frame dilaton ($\varphi$) is simply related to $\phi$ as 
$\varphi = \phi\sqrt{2/(n + 2)}$.
The transformation of Eq. (\ref{transf2}) 
 clearly induces a change in the 
curvature invariants. Consider, for instance, the scalar curvature in the string 
frame and call it $R$. In the Einstein frame the scalar curvature 
will be $ {\cal R}$. The relation between $R$ and ${\cal R}$ is 
\begin{equation}
R = e^{- \psi} \biggl[ {\cal R} - ( n + 3) G^{\alpha\beta} \nabla_{\alpha} \nabla_{\beta} \psi 
- \frac{(n + 3) (n + 2)}{4} G^{\alpha\beta} 
\nabla_{\alpha}\psi \nabla_{\beta}\psi\biggr],
\label{trascur}
\end{equation}
where $\psi =  \varphi\sqrt{2/( n + 2)}$ 
and where the covariant derivatives are 
computed with respect to $G_{\mu\nu}$.
The two frames certainly 
describe the same physics even if there has been debate on which of the 
two frames captures better the stringy features of a given solution of the low-energy 
$\beta$-functions. As far as the singularity properties are concerned we can clearly 
see that the two frames are not equivalent. Suppose, for instance, that 
the dilaton field is decreasing in the Einstein frame and suppose that ${\cal R}$ 
is regular in the same frame. Then Eq. (\ref{trascur}) tells that 
$R$ will be no longer regular in the string frame. Therefore 
singularities may appear and disappear from one frame to the other.

This example only means that the two frames are indeed 
physically equivalent when we deal with {\em physical } solutions 
namely solutions where the dilaton is asymptotically constant, for instance. 
This observation helps in setting the goals of the present paper. 
We want warped compactifications, corresponding to stable critical points of 
the truncated $\beta$-functions for which the dilaton field is frozen to a 
constant value. 

A similar situation \cite{mgst} occurs in another type of 
inhomogeneous backgrounds (i.e. four-dimensional background geometries 
with two commuting Killing vectors which are hypersurface 
orthogonal and orthogonal to each others). In \cite{mgst} completely regular and 
geodesically complete solutions of the low-energy $\beta$ functions
have been discussed in the framework of these backgrounds. 
Also in the case \cite{mgst} the crucial 
requirement is that the dilaton interpolates between two constant 
values.
\renewcommand{\theequation}{B.\arabic{equation}}
\setcounter{equation}{0}
\section{Equations of motion with the first $\alpha'$ correction}
Inserting the metric given in Eq. (\ref{line}) into Eq. (\ref{action2}) 
 the reduced form of the action is obtained 
to first order in $\alpha'$, namely
\begin{eqnarray}
&&S= - \frac{1}{2 \lambda_{s}^4}\int d^4 x\, \int d \rho\,\, d\theta\,\, 
\,\,\sigma^2\,\, \sqrt{\gamma}\,\, e^{- \phi}\,\, 
\{\frac{\F^2}{2} + 5 {\cal H}^2 + 2 {\cal H}{\cal F} 
+ 4 \partial_{\rho} {\cal H} 
+ \partial_{\rho} {\cal F} - (\partial_{\rho}\phi)^2 + \Lambda 
\nonumber\\
&&- \epsilon [3 {\cal H}^2 {\cal F}^2 + 
\frac{15}{2} {\cal H}^4 + 6 {\cal H}^2 \partial_{\rho} {\cal F} 
+ 12 {\cal H}{\cal F} \partial_{\rho} {\cal H} + 12 {\cal H}^3 {\cal F} 
+ 12 {\cal H}^2 
\partial_{\rho} {\cal H} - (\partial_{\rho} \phi)^4 ]\}.
\end{eqnarray}
Recalling now that ${\cal H} = \partial_{\rho} \ln{\sigma}$ and that 
${\cal F} = \partial_{\rho}
\ln{\gamma}$ we can perform, separately, the variation with 
respect to $\phi$, $\sigma$ and $\gamma$. 
The variation with respect to $\phi$ gives
\begin{eqnarray}
&&-2\,{\L} - {\F}^2 - 4\,\F\,\H - 10\,{\H}^2 - 2\,\F' - 8\,\H' + 
  2\,\F\,{\phi}' + 8\,\H\,{\phi}' - 2\,{{\phi}'}^2 + 
  4\,{\phi}'' 
\nonumber\\
&& + \epsilon\,( 6\,{\F}^2\,{\H}^2 + 24\,\F\,{\H}^3 + 15\,{\H}^4 + 
     12\,{\H}^2\,\F' + 24\,\F\,\H\,\H' + 24\,{\H}^2\,\H' 
\nonumber\\
&& - 4\,\F\,{{\phi}'}^3 - 
     16\,\H\,{{\phi}'}^3 + 6\,{{\phi}'}^4 - 
     24\,{{\phi}'}^2\,{\phi}'' ) =0,
\label{trea}
\end{eqnarray}
whereas the variation with respect to $\sigma$ and $\gamma$ gives 
\begin{eqnarray}
&& f_{3}''(\rho) + 2 ({\cal H} + \frac{{\cal F}}{2} - \phi') f_{3}'(\rho) 
- f_{2}'(\rho) - 
({\cal H} + \frac{{\cal F}}{2}  - \phi') f_{2}(\rho) 
\nonumber\\
&&+ 
\biggl[ ({\cal H} + \frac{{\cal F}}{2} - \phi')^2 +   ({\cal H}' 
+ \frac{{\cal F}'}{2} - \phi'')\biggr]
f_{3}(\rho) + f_1(\rho) =0,
\label{unoa}
\end{eqnarray}
\begin{eqnarray}
&& g_{3}''(\rho) + 2 (2{\cal H} - \frac{{\cal F}}{2} - \phi') g_{3}'(\rho) 
- g_{2}'(\rho) - 
(2{\cal H} - \frac{{\cal F}}{2}  - \phi') g_{2}(\rho) 
\nonumber\\
&&+ 
\biggl[ (2{\cal H} - \frac{{\cal F}}{2} - \phi')^2 
+   (2{\cal H}' -\frac{{\cal F}'}{2} - \phi'')\biggr]
g_{3}(\rho) + g_1(\rho) =0
\label{duea}
\end{eqnarray}
where the $f(\rho)$ are given by
\begin{eqnarray}
&&f_{1}(\r) = 2\,\L - 2\,\F\,\H + {\left( \F + 2\,\H \right) }^2 + 
  2\,\left( \F' + 2\,\H' \right)  - 2\,{{\phi}'}^2 + 
  \epsilon\,\left( 3\,{\H}^4 + 12\,{\H}^2\,\H' + 2\,{{\phi}'}^4 \right),
\nonumber\\
&&f_2(\r) = 2\,\left( \F + \H \right)  - 6\,\epsilon\,\left[ {\H}^3 
+ \F\,\H\,\left( 2\,\H + \F\right)  + 
     4\,\H\,\H' + 2\,\left( \H\,\F' + \F\,\H' \right)  \right],
\nonumber\\
&& f_3(\r) = 4[1 - 3\,\epsilon\,\H\,\left( \F + \H \right)],
\label{f}
\end{eqnarray}
and the $g(\r)$ are given by
\begin{eqnarray}
&&g_1(\r) =\frac{1}{4}\biggl[2\,{\L} + {\F}^2
 - 4\,\F\,\H + 10\,{\H}^2 - 2\,\F' + 8\,\H' - 
    2\,{{\phi}'}^2, 
\nonumber\\
&&- \epsilon\,\left( 6\,{\F}^2\,{\H}^2 - 24\,\F\,{\H}^3 + 15\,{\H}^4 - 
       12\,{\H}^2\,\F' - 24\,\F\,\H\,\H' + 24\,{\H}^2\,\H' 
- 2\,{{\phi}'}^4 \right)\biggr],
\nonumber\\
&& g_2(\r) = 
-\F + 2\,\H - 6\,\epsilon\,\left( - \F\,{\H}^2   + 2\,{\H}^3 + 2\,\H\,\H' 
\right),
\nonumber\\
&& g_3(\r) = 1 - 6\,\epsilon\,{\H}^2.
\label{g}
\end{eqnarray}
Recall that $' = \partial_{\rho}$.
Eq. (\ref{trea}) exactly coincides with Eq. (\ref{tre}). Inserting 
Eqs. (\ref{f})--(\ref{g}) 
into Eqs. (\ref{unoa}) and (\ref{trea}) we get, respectively, 
Eqs. (\ref{uno}) and (\ref{due}).

\renewcommand{\theequation}{C.\arabic{equation}}
\setcounter{equation}{0}
\section{Eigenvalues around the critical points}
The two eigenvalues in the case of $\L <0$ can be written as 
\begin{eqnarray}
&& W_1(k) = \sqrt{\frac{5\,\L_{-}}{2}}
{\sqrt{1 + {\sqrt{1 + \frac{212\,k}{5}}}}} \frac{\biggl( T(k) 
+ \sqrt{ S(k) Q(k)} \biggr)}{V(k)}
\label{w1}\\
&& W_2(k) =  \sqrt{\frac{5\,\L_{-}}{2}}
{\sqrt{1 + {\sqrt{1 + \frac{212\,k}{5}}}}}
 \frac{\biggl( T(k) - \sqrt{ S(k) Q(k)} \biggr)}{V(k)}
\label{w2}
\end{eqnarray}
where
\begin{eqnarray}
&& T(k) =\frac{1}{2}
\biggl\{ -1925 - 260180\,k - 7570944\,k^2 + 
      5\,{\sqrt{1 + \frac{212\,k}{5}}}\,[ 385 + 384\,k\,
\left( -19 + 792\,k \right) ] \biggr\} 
\nonumber\\
&& S(k) =\biggl\{ 1015 + 99916\,k + 2411712\,k^2 - 
      {\sqrt{1 + \frac{212\,k}{5}}}\,[ 1015 + 12\,k\,\left( 2269 
+ 158400\,k \right) ]  \biggr\} 
\nonumber\\
&& Q(k) =\biggl\{ 125 + 123380\,k + 5006592\,k^2 + 
      5\,{\sqrt{1 + \frac{212\,k}{5}}}\,[ -25 + 48\,k\,\left( 563 
+ 12672\,k \right)  ]  \biggr\} 
\nonumber\\
&& V(k) =  \,k\,[ 127925 + 576\,k\,\left( 12473 + 348480\,k \right)] 
\end{eqnarray}
Recall that $k = \epsilon \L_{-}$.

\end{appendix}

\end{document}